\documentclass{article}
\usepackage{arxiv}
\usepackage[utf8]{inputenc} 
\usepackage[T1]{fontenc}    
\pdfoutput=1
\usepackage{natbib}
\usepackage{hyperref}       
\usepackage{url}            
\usepackage{booktabs}       
\usepackage{amsfonts}       
\usepackage{nicefrac}       
\usepackage{microtype}      
\usepackage{amsmath,amsthm,amssymb}
\usepackage{graphicx}
\usepackage{doi}
\usepackage[table,xcdraw]{xcolor}
\usepackage{xcolor}

\usepackage{color}
\definecolor{purple}{rgb}{0.459,0.109,0.538}

\newcommand{\id}[1]{\textcolor{brown}{*** ID: #1}}

\newcommand{\comment}[1]{}

\title{A Surrogate Endpoint Based Provisional Approval Causal Roadmap}

\author{ \href{https://orcid.org/0000-0002-2662-9427}{\includegraphics[scale=0.06]{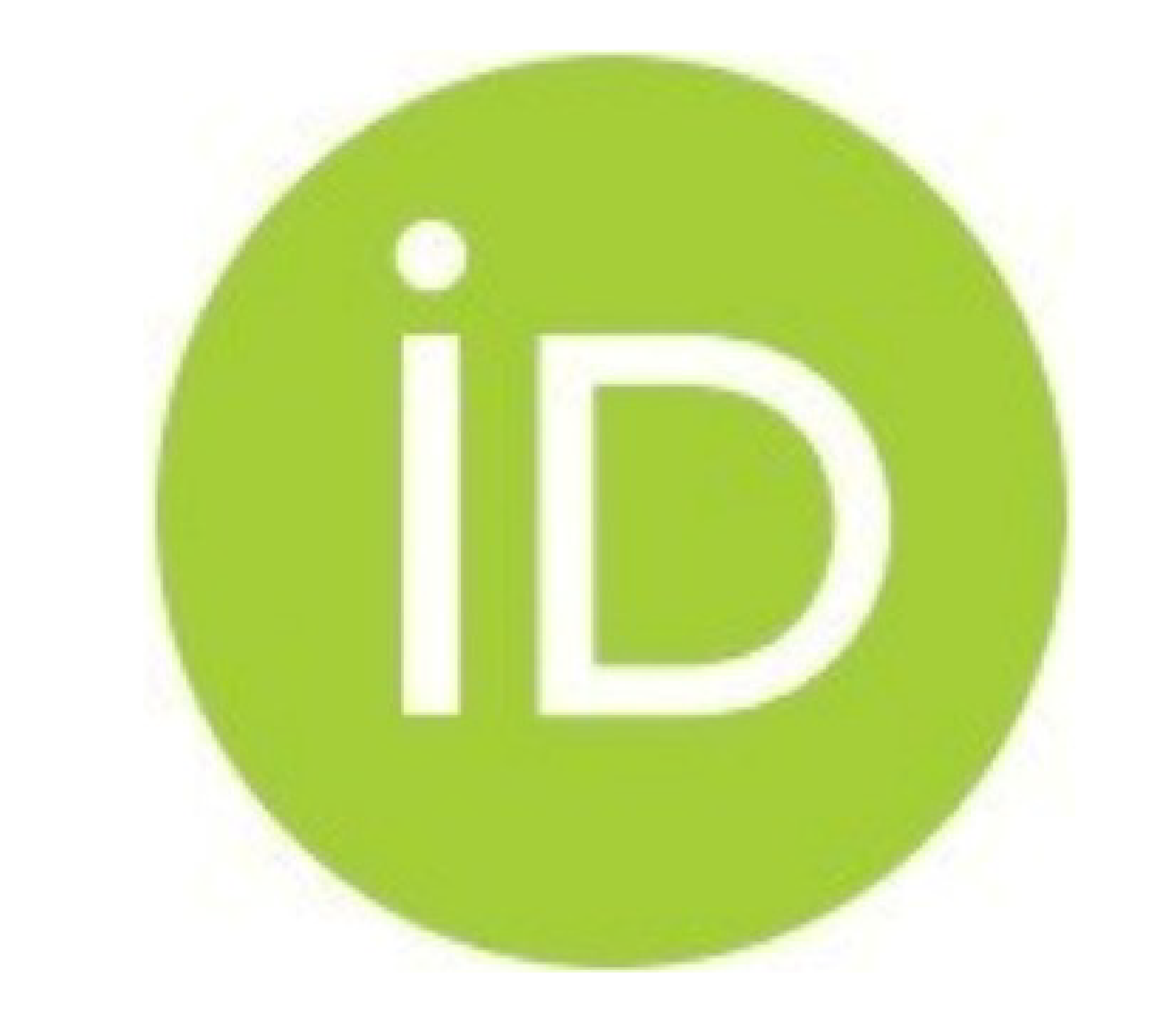}\hspace{1mm}Peter B.~Gilbert} \\
	Vaccine and Infectious Disease and Public Health Sciences Divisions\\
	Fred Hutchinson Cancer Center\\
	Seattle, WA 98109 
        Department of Biostatistics \\
        University of Washington \\
        Seattle, WA 98105 \\
	\texttt{pgilbert@fredhutch.org} \\
	\And
	\href{https://orcid.org/0000-0002-8245-8319}{\includegraphics[scale=0.06]{orcid.png}\hspace{1mm}James~Peng} \\
	Department of Biostatistics\\
	University of Washington\\
	Seattle, WA 98105 \\
	\texttt{jpspeng@uw.edu} \\
 \And
	\href{https://orcid.org/0000-0002-0577-9661}{\includegraphics[scale=0.06]{orcid.png}\hspace{1mm}Larry~Han} \\
	Department of Public Health and Health Sciences\\
	Bouv\'{e} College of Health Sciences \\
	Northeastern University, MA 02115 \\
	\texttt{lar.han@northeastern.edu} \\
 \And
\href{https://orcid.org/0000-0001-6807-8347}{\includegraphics[scale=0.06]{orcid.png}\hspace{1mm}Theis~Lange} \\
Department of Public Health \\
University of Copenhagen \\
Copenhagen, K-1353, Denmark \\
	\texttt{thlan@sund.ku.dk} \\
 \And
	\href{https://orcid.org/0000-0001-9332-6832}{\includegraphics[scale=0.06]{orcid.png}\hspace{1mm}Yun~Lu} \\
	Center for Biologics Evaluation and Research \\
	Food and Drug Administration \\
	Silver Spring, MD 20993
 \\
	\texttt{yun.lu@fda.hhs.gov} \\
\And
	\href{https://orcid.org/0000-0002-7791-9395}{\includegraphics[scale=0.06]{orcid.png}\hspace{1mm}Lei~Nie} \\
	OB/OTS/CDER \\
	Food and Drug Administration \\
	Silver Spring, MD 20993 \\
	\texttt{lei.nie@fda.hhs.gov} \\
\And
	\href{https://orcid.org/0000-0002-3924-5668}{\includegraphics[scale=0.06]{orcid.png}\hspace{1mm}Mei-Chiung~Shih} \\
	VA Palo Alto Health Care System \\
	795 Willow Road \\
	Menlo Park, CA 94025 \\
	\texttt{mei-chiung.shih@va.gov} \\
\And
	\href{https://orcid.org/0000-0000-0000-0000}{\includegraphics[scale=0.06]{orcid.png}\hspace{1mm}Salina P.~Waddy} \\
	Division of Clinical Innovation\\
    National Center for Advancing Translational Sciences\\
    National Institutes of Health \\
    Bethesda, Maryland 20892 \\ 
	\texttt{salina.waddy@nih.gov} \\
\And
	\href{https://orcid.org/0000-0000-0000-0000}{\includegraphics[scale=0.06]{orcid.png}\hspace{1mm}Ken~Wiley} \\
	Division of Clinical Innovation\\
    National Center for Advancing Translational Sciences\\
    National Institutes of Health \\
    Bethesda, Maryland 20892 \\ 
	\texttt{ken.wiley@nih.gov} \\
\And
	\href{https://orcid.org/0009-0004-9232-0382}{\includegraphics[scale=0.06]{orcid.png}\hspace{1mm}Margot~Yann} \\
	School of Public Health, UC Berkeley \\
	Forum for Collaborative Research, UCDC Campus \\
	Washington, DC, 20036 \\
	\texttt{margot.yann@berkeley.edu} \\  
 \And
	\href{https://orcid.org/0000-0003-2507-4949}{\includegraphics[scale=0.06]{orcid.png}\hspace{1mm}Zafar~Zafari} \\
	University of Maryland School of Pharmacy \\
        Baltimore, MD 21201
	University of Maryland Institute for Health Computing \\
	Bethesda, MD 20852 \\
	\texttt{zzafari@rx.umaryland.edu} \\  
 \And
	\href{https://orcid.org/0000-0001-6618-1316}{\includegraphics[scale=0.06]{orcid.png}\hspace{1mm}Debashis~Ghosh} \\
	Department of Biostatistics $\&$ Informatics \\
	Colorado School of Public Health \\
	Aurora, CO 80045 \\
	\texttt{debashis.ghosh@cuanschutz.edu} \\  
 \And
	\href{https://orcid.org/0000-0003-4073-0393}{\includegraphics[scale=0.06]{orcid.png}\hspace{1mm}Dean~Follmann} \\
	Biostatistics Research Branch \\
	National Institute of Allergy and Infectious Diseases \\
    Rockville, MD 20852 \\
	\texttt{dfollmann@niaid.nih.gov} \\  
 \And
	\href{https://orcid.org/0000-0002-0920-2915}{\includegraphics[scale=0.06]{orcid.png}\hspace{1mm}Michal~Juraska} \\
	 Vaccine and Infectious Disease Division\\
	Fred Hutchinson Cancer Center\\
	Seattle, WA 98109 \\
	\texttt{mjuraska@fredhutch.org} \\
 \And
	\href{https://orcid.org/0000-0001-9056-2047}{\includegraphics[scale=0.06]{orcid.png}\hspace{1mm}Iv\'{a}n~D\'{i}az} \\
	 Department of Biostatistics \\
	New York University Grossman School of Medicine \\
	New York, NY 10016 \\
	\texttt{ivan.diaz@nyu.edu} \\
}

\renewcommand{\shorttitle}{Provisional Approval Causal Roadmap}

\hypersetup{
pdftitle={A Surrogate Endpoint Based Provisional Approval Causal Roadmap},
pdfauthor={Peter B.~Gilbert, James~Peng, Larry~Han, Theis~Lange, Yun~Lu, Lei~Nie, Mei-Chiung~Shih, Salina P.~Waddy, Ken~Wiley, Margot~Yann, Zafar~Zafari, Debashis~Ghosh, Dean~Follmann, Michal~Juraska, Iv\'{a}n~D\'{i}az},
pdfkeywords={Causal inference, Causal roadmap, Group B Streptococcus, Sensitivity analysis, Surrogate endpoint, Transportability},
}

\begin{document}
\maketitle

\begin{abstract}
For many rare diseases with no approved preventive interventions,  
promising interventions exist, yet it has been difficult to conduct a pivotal phase 3 trial that could provide direct evidence demonstrating a beneficial effect on the target disease outcome. 
When a promising putative surrogate endpoint(s) for the target outcome
is available, surrogate-based provisional approval of an intervention may be pursued.
We apply the Causal Roadmap rubric to define a surrogate endpoint based provisional approval causal roadmap, which combines observational study data that estimates the relationship between the putative surrogate and the target outcome, with a phase 3 surrogate endpoint study that collects the same data but 
is very under-powered to assess the treatment effect (TE) on the target outcome.
The objective is conservative estimation/inference for the TE with an estimated lower uncertainty bound that 
allows (through two bias functions) for an imperfect surrogate and imperfect transport of the conditional target outcome risk in the untreated between the observational and phase 3 studies. Two estimators of TE (plug-in, nonparametric efficient one-step) with corresponding inference procedures are developed. Finite-sample performance of the plug-in estimator is evaluated in two simulation studies, with R code provided.
The roadmap is illustrated with contemporary Group B Streptococcus vaccine development.
\end{abstract}


\section{Introduction}

For more than 10,000 rare diseases, no effective treatments are licensed/approved \citep{fermaglich2023comprehensive}. 
    The traditional pathway for approving treatments generates evidence of treatment effectiveness based on one or preferably two randomized, controlled phase 3 trials that directly demonstrate benefit on a target outcome of interest that reflects how an individual ``feels, functions, or survives" \citep{FlemingDeMets1996,temple1999surrogate,FlemingPowers2012}. When the target outcome is defined by the onset of a rare disease, i.e., for treatments aiming to prevent disease onset, the outcome event itself is rare. Because the target outcome is rare, the phase 3 trial would need a very large sample size (i.e., hundreds of thousands) to be powered to demonstrate benefit.  It is frequently very challenging to garner enough resources to conduct such phase 3 trials.  


Yet, promising candidate surrogate endpoints are sometimes available that open an alternative approval pathway to pursue: provisional approval based on phase 3 trials that use the surrogate endpoint as the primary endpoint. 
By ``surrogate endpoint" we use the definition from a 2016 NIH/FDA
workshop \citep{food2016best} that was voted by the SPIRIT-SURROGATE/CONSORT-SURROGATE project team as a preferred definition \citep{ciani2023framework}: paired to a target outcome of interest, a surrogate endpoint is an intermediate outcome that itself does not reflect ``feels, functions, or survives" but can be used as a substitute for the target outcome to reliably provide estimation and inference for the treatment effect on this target outcome.  
The U.S. FDA's accelerated approval regulation codifies this pathway \citep{fda1992accelerated,fda2021}, which enables provisional approval of a treatment for a serious or life-threatening disease with unmet need based on a sufficiently well validated surrogate endpoint for a target outcome.  This pathway requires a commitment to post-approval studies that would directly verify the effectiveness of the treatment against the target outcome.

Our motivating example is development of vaccines
against Group B Streptococcus (GBS), which causes invasive GBS disease (IGbsD) in infants and is a leading cause of young infant death \citep{gonccalves2022group}. No vaccine has been approved to prevent young infant IGbsD.  The World Health Organization (WHO) has identified development of a GBS vaccine for immunization during pregnancy as a priority  \citep{kobayashi2016group}. While multiple companies are developing maternal GBS vaccines \citep{vekemans2019role}, no phase 3 vaccine efficacy trial has been conducted, partly because the trial would need to be very large given the low incidence of IGbsD of about 1--3 per 1000 live births  \citep{vekemans2019maternal}. Therefore, the GBS vaccine field is currently pursuing a provisional approval pathway based on antibody markers measured in cord blood that have been shown to strongly inversely correlate with IGbsD in natural history studies
[e.g., \citet{madhi2021association,madhi2023potential,dangor2023association}].
The US FDA's Vaccine Advisory Board in May of 2018 concluded that 
one of these antibody markers, which measures 
concentration of IgG antibodies binding to specific GBS proteins, is reasonably likely to predict vaccine efficacy of capsular polysaccharide GBS vaccines against IGbsD, and recommended a provisional approval pathway based on this surrogate endpoint \citep{gilbert2022methodology}.
Moreover, the EMA Guideline on clinical evaluation of vaccines EMEA/CHMP/VWP/164653/05 Rev. 1 mentioned that if a vaccine efficacy trial is not feasible, then an indication of the immune parameter of greatest importance for protection and sometimes a preliminary CoP may be obtained from one or more sero-epidemiological studies (i.e., examining natural protection). 

The Causal Roadmap is a general seven-step framework for pursuing answers to causal questions \citep{petersen2014causal,dang2023causal}. 
In this article, we apply the Causal Roadmap to outline a recipe for generating evidence for supporting a provisional approval decision for the context noted above, using contemporary GBS vaccine development as a running example. Our scope considers contexts meeting all three of the following conditions: (1) no randomized, controlled phase 3 trials have been conducted of the candidate treatment of interest and it has proven formidable to conduct such phase 3 trials such that none are expected on the horizon, (2) one or more prospective observational studies have been conducted in untreated persons that assess the relationship of one or more candidate surrogate endpoints with a target outcome of interest, 
and (3) a treatment for the disease is being developed with a provisional approval strategy via a pivotal randomized, controlled phase 3 trial with the candidate surrogate endpoint as the primary endpoint. 

\comment{
{\color{blue} PG: 11-24-23: For Step 3, currently this manuscript considers a prospective cohort observational study, without the augmentation of additional cases.  As noted below, developing this work to allow an augmented part is important, so if a co-author gets inspired to contribute this hybrid elaboration in this current manuscript that would be neat.  

In addition, it would be interesting to consider a case-control observational study without a prospective cohort component; perhaps this would make sense as a companion separate manuscript given this one is already getting long. }
}

In addition to this phase 3 trial being an actual trial that is well-powered to study the treatment effect on the surrogate endpoint, we envisage a hypothetical version of this
phase 3 trial -- a hypothetical ``target trial" \citep{hernan2016using, hernan2022target} --
for which we imagine enrolling a vastly larger sample size that would power the trial to assess the treatment effect on the target outcome.  
Based on a harmonized data set from an observational study (or studies)
to learn a surrogate and from a phase 3 trial to apply the surrogate,
a statistical approach is needed to estimate the treatment effect on the target outcome in the target trial with an estimated uncertainty interval (EUI) codified in some fashion.  We consider that the primary success criterion for provisional approval may be defined by this EUI lying above a pre-specified minimum lower bound of the treatment effect on the target outcome determined through iterative deliberations with regulators, and may also require a minimal lower bound for the point estimate of the treatment effect.

 As such, 
our objective is to apply the Causal Roadmap to develop a process, culminating in a statistical analysis plan encompassing the observational and phase 3 studies, for transparently obtaining and interpreting the EUI and point estimates that determine success or failure for meeting provisional approval criteria. (Regulators would account for additional information that cannot all be codified into the statistical analysis plan.)
Current GBS vaccine development research
 fits the above scope, serving as our illustrative example.  Indeed, multiple vaccine developers are pursuing this approach that characterizes an antibody surrogate endpoint in multiple sero-epidemiological natural history observational studies and then conducts a phase 3 surrogate endpoint trial to evaluate qualification against provisional approval success criteria. 
Envisaged as a target trial, the primary objective of the phase 3 trial is estimation/inference about vaccine efficacy against IGbsD for the phase 3 study population based on the surrogate defined from data analysis of the observational studies.  Our application of the Causal Roadmap provides a way to statistically answer this primary objective.
While confirmation of benefit in post-provisional-approval validation studies is crucial, it is out of our scope to apply the Causal Roadmap for these studies; \citet{dang2023causal} describes the Causal Roadmap for generating high-quality real-world evidence that applies for this purpose.  

The problem we address is extending (generalizing or transporting) causal inferences from one study to a new target population that is different from the original study population.  The literature has focused
on extending causal inferences from one or more randomized trials to a target population represented by an observational study sample (e.g., \cite{ColeStuart2010,westreich2017transportability,buchanan2018generalizing,li2023efficient}). Particularly germane to our problem, \citet{rudolph2017robust}, \citet{dahabreh2020extending}, and \citet{dahabreh2023sensitivity}
developed methods including robust targeted maximum likelihood, g-formula, inverse probability weighting, and combined double-robust methods for extending causal inferences about a point treatment effect from a randomized trial to a target population of nonparticipants, based on all the data from the randomized trial and 
a sample of baseline covariates from the target population, where the utility of the baseline covariates is to correct for bias from treatment effect modifiers influencing participation in the randomized trial. 
\citet{dahabreh2023sensitivity} is especially germane because they developed methods for sensitivity analysis via bias functions that provide conservative lower-bound inferences for the treatment effect in the target population, fitting the essential requirement of our provisional approval objective.  However, this work departs from \citet{dahabreh2023sensitivity} in two main ways. First, in our problem context, knowledge learned from an observational study (with no treatment) is applied to make inferences about a treatment effect in a new target population that is studied in a randomized trial (reversing the role of the randomized and observational study). Second, in our problem knowledge is extended based on the relationship between an intermediate response surrogate endpoint and baseline covariates on the target outcome, instead of only considering baseline covariates. The implication is that different causal identifiability assumptions are needed to license valid inferences on the causal treatment effect in the target population.  Another implication is that the notion of a valid surrogate endpoint is important for our problem, where there is a large literature on surrogate endpoint evaluation based on one or multiple target outcome randomized trials (e.g., \cite{alonso2015relationship}). However, because in our scenario no phase 3 target outcome randomized trial has yet been done, evidence for the appropriateness of the surrogate will need to come from other sources, and in our development below we describe the requirements for the surrogate endpoint to yield the desired correct inferences.


As a surprise to us, just before we were ready to post this work on arXiv, we discovered (on May 29, 2024) the pre-print
\citet{athey2024surrogate} that addressed essentially the same statistical problem.
\citet{athey2024surrogate} considered inference on the same treatment effect target causal parameter of interest, based on the same collected data with one difference that our work assumes all observational study participants have treatment known to be the control condition whereas Athey et al. assumes  observational study participants have treatment missing/unknown. Under both a comparability assumption and a perfect surrogate assumption (A4 and A6 below), the identifiability results and nonparametric efficient influence function are equivalent in the two articles, whereas with an imperfect surrogate, the results differ as a consequence of the different set-up. There are several differences in the articles in the detailed methodologies that are developed, including different versions of the comparability and perfect surrogate assumptions, the present article focuses on accommodating missing data on the surrogate(s) and considers other practical challenges such as right-censoring of the target outcome and intercurrent events, the present article focuses on a general contrast treatment effect parameter compared to Athey et al. that focuses on an additive difference contrast, and the present article focuses on conservative inference throughout as this is prioritized for the provisional approval application. The present article considers many issues specifically germane to the provisional approval application not considered by Athey et al. As general statistical methodology, we consider the two articles as independent works that provide similar and in many ways equivalent results, emanating from different fields -- \citet{athey2024surrogate} from economics and ours from biomedical clinical sciences. Section \ref{diffarticles} summarizes equivalencies, similarities, and differences between the two articles.

\section{Notation and Set-Up of the Data Sources: Observational Study and Phase 3 Study}
\label{notation}


Our notation is similar to 
\citet{dahabreh2023sensitivity}, except using $Z$ instead of $S$ to denote study.  
%
We consider a single observational study harmonized with a single phase 3 trial.  Let $Z=1$ ($Z=0$) indicate enrollment into the observational (phase 3) study.  We enroll $n_{obs}$ individuals into the observational study, with covariates $X$ measured at enrollment and intermediate outcomes $S$ measured after enrollment and
by the fixed visit $\tau$ post-enrollment (in many applications $S$ is biomarkers measured from a blood sample drawn at time $\tau$). 
Participants are followed after $\tau$ over a fixed period through time $t_0$ for whether the target outcome occurs: $Y=I(T \le t_0)$ where $T$ is the time from $\tau$ to the target outcome failure event and $I(B)$ is the indicator of an arbitrary event $B$; also let $C$ be the time from $\tau$ until right-censoring, with $\tilde T = min(T,C)$ and $\Delta$ the indicator of observed failure after $\tau$ by $t_0$, $\Delta = I(T \le C)$. Follow-up through $t_0$ without the event means $\tilde T = C$ and $\Delta = 0$. In many applications, $S$ is only meaningfully defined if the participant did not experience the target outcome by $\tau$, in which case such participants must be excluded for the purpose of surrogate endpoint evaluation. With $Y^{0-\tau}=1$ the indicator of target outcome occurrence after enrollment by time $\tau$, in such applications only participants with $Y^{0-\tau}=0$ are eligible for inclusion. Note that failure by $\tau$ is a distinct issue from loss to follow-up by $\tau$.  While participants lost to follow-up by $\tau$ are also excluded, the parameters of interest are defined for a hypothetical world with no loss to follow-up, whereas in contrast for such applications this hypothetical world is not considered conceptually viable for early failure. That is, early right-censoring is a nuisance, with $S$ still well-defined for the vast majority of participants who will not fail by $\tau$ (with complication this is not measurable), whereas early-failure renders $S$ undefined. Section \ref{ICEsec} discusses this inter-current event issue further. 

For study design, 
$\tau$ is selected for viability of defining a surrogate endpoint based on measurements of $S$ up to and including time $\tau$, where broad inter-individual variability in $S$ across participants in the $Z=0$ and $Z=1$ studies is a factor that improves precision for estimation of the treatment effect parameter.  It is desirable to select $\tau$ to be fairly close to enrollment, to attain the practical advantage of a surrogate endpoint to facilitate shorter studies. For vaccine studies, $\tau$ is typically defined as a study visit 2--4 weeks after primary vaccination, a visit at which antibody markers have wide inter-vaccinee variability.

Given that the target outcome $Y=1$ is rare, for resource efficiency $S$ is measured in a random sample of participants
 with $Y^{0-\tau}=0$ such as through two-phase case-cohort sampling.  Let $\epsilon_S$ be the indicator that $S$ is measured.
Let $A=0$ indicate that a participant does not receive the treatment being developed; while all participants in the observational study are not treated, we include this notation to emphasize this basic feature of the study.  

In the phase 3 trial, participants are randomized to receive the treatment $A=1$ or a control condition $A=0$ such as placebo that domain knowledge supports should have the same meaning as $A=0$ in the observational study in terms of not having an impact on $Y$. If $A=0$ in the phase 3 trial is placebo then this condition will hold (truism of a valid placebo), but if $A=0$ is an active control then assurance is needed that had, hypothetically, the active control been used by all observational study participants it would not have impacted risk of $Y$ compared to receiving nothing. 
 The same variables $X$, $Y^{0-\tau}$, $\epsilon_S$, $\epsilon_S S$, $\tilde T$, $\Delta$ measured in the observational study are measured, where again the intermediate outcome(s) $S$ may be only measured in a random sample.  The notation $\epsilon_S S$ means that $S$ is observed if $\epsilon_S=1$ and not observed if 
 $\epsilon_S=0$.

To simplify exposition, throughout we assume
all enrolled participants in both studies attend the visit $\tau$ at which $S$ is measured without experiencing the target outcome, i.e., $Y^{0-\tau}_i = 0$ for all $i$. 
In the discussion section we consider potential extensions relaxing this assumption.  In total, the composite data set consists of 
 $n = n_{obs} + n_{RCT}$ observations, with $n_{obs}$ iid observations 
 $(X_i,Z_i=1,A_i=0,\epsilon_{Si}, \epsilon_{Si} S_i, \tilde T_i, \Delta_i)$
 and 
 $n_{RCT}$ iid observations 
$(X_i,Z_i=0,A_i,\epsilon_{Si},\epsilon_{Si} S_i)$,
where $Y_i = I(T \le t_0)$ has a known value if
$\tilde T = t_0$ and $\Delta=0$ or if $\tilde T \le t_0$ and $\Delta=1$.
In some applications $\tilde T_i$ and $\Delta_i$ are also collected in the phase 3 study; yet
 an essential feature of our problem set-up is very few observed $Y_i=1$ events are expected in the phase 3 study (less than 10 in phase 3 GBS vaccine studies), and all our results (identifiability and estimation) do not make use of any  
 $\tilde T_i$ and $\Delta_i$ values of phase 3 participants.

The relevant $S$ are intermediate outcomes that, based on domain knowledge, can potentially predict $Y=1$ and may be connected to treatment-efficacy mechanisms and hence possibly contribute to a surrogate endpoint.  In some applications
$S$ could be a single intermediate outcome, whereas in others it may be a multivariate set of intermediate outcomes. 
Regardless of the dimensionality of $S$, we consider the 
function 
$$g(x,s) := P(Y=1|X=x,Z=1,A=0,S=s)$$ 

\noindent to be the central ingredient for estimating treatment efficacy in the 
phase 3 study, where the observational study is used to
develop an optimal estimator for $g(X,S)$. 
 \citet{PriceGilbertvanderLaan2018} proposed that 
 an optimal estimator of $g(X,S)$ -- a so-called
 estimated optimal surrogate (EOS) -- be considered for use as a surrogate endpoint. The EOS has the useful feature that it is on the scale of the absolute risk of the target outcome of interest, aiding its biomedically-relevant interpretation.  The data analysis of the observational study for estimation of $g(X,S)$ may consider many different input variable sets $(X,S)$ and different ways of entering the variables into models, seeking empirical learning of a most promising EOS.
A selected EOS from the observational study is used in the estimators of treatment efficacy in the phase 3 study. Alternative optimal transformations of the surrogate have been proposed to directly target the treatment effect on the target outcome when information on both the treatment and control conditions is available \citep{han2022identifying}. 

In the phase 3 study, it is not required to measure all the components of $X$ and $S$ that were measured in the observational study; it is only required to measure the components that
are used in the estimator of $g(X,S)$ in the observational study.
  Therefore empirical learning of $g(X,S)$ in the observational study could be used to winnow down to a subset of $(X,S)$ variables to measure in the phase 3 study to potentially save resources and focus on a more parsimonious surrogate endpoint. Justification for doing this would include learning that not all $(X,S)$ are needed for obtaining an EOS, and evidence that the selected $(X,S)$ are sufficient for meeting the causal assumptions (listed in Step 4) and not widening estimated uncertainty intervals.  
  For vaccine applications, we suspect that applications where 
$g(X,S)$ can be defined based on one- or two-variate $S$ will be most successful. 


\section{Application of the Causal Roadmap Framework to Surrogate-Endpoint Based Provisional Approval}

\comment{
Aim to develop it agnostically to the tools/methods used to derive causal inferences.
The Causal Roadmap is a framework for answering causal questions (ref). It focuses on delineating the steps and assumptions necessary to make causal inferences or answer causal questions. The steps in the roadmap are agnostic to the tools/methods used to derive causal inferences.
The seven steps in the general roadmap for causal inference are:

\begin{enumerate}
\item Specify knowledge about the system to be studied using a causal model: What do we already know?
\item Define the causal parameter of interest.
\item Translate the causal parameter into a statistical estimand.
\item Identify conditions under which the statistical estimand equals or bounds the causal parameter. 
\item Estimate the statistical estimand.
\item Quantify the uncertainty in the estimate of the statistical estimand.
\item Interpret the estimate of the causal parameter.
\end{enumerate}
}

In this section, we apply each step of the Causal Roadmap to the surrogate endpoint provisional approval application.  
The subsequent section applies these steps specifically to the GBS vaccine case study.

\subsection{Step 1: Specify the causal model based on available knowledge of the context and proposed study}

\begin{figure}[h!]
    \centering
    \includegraphics[width=\textwidth]{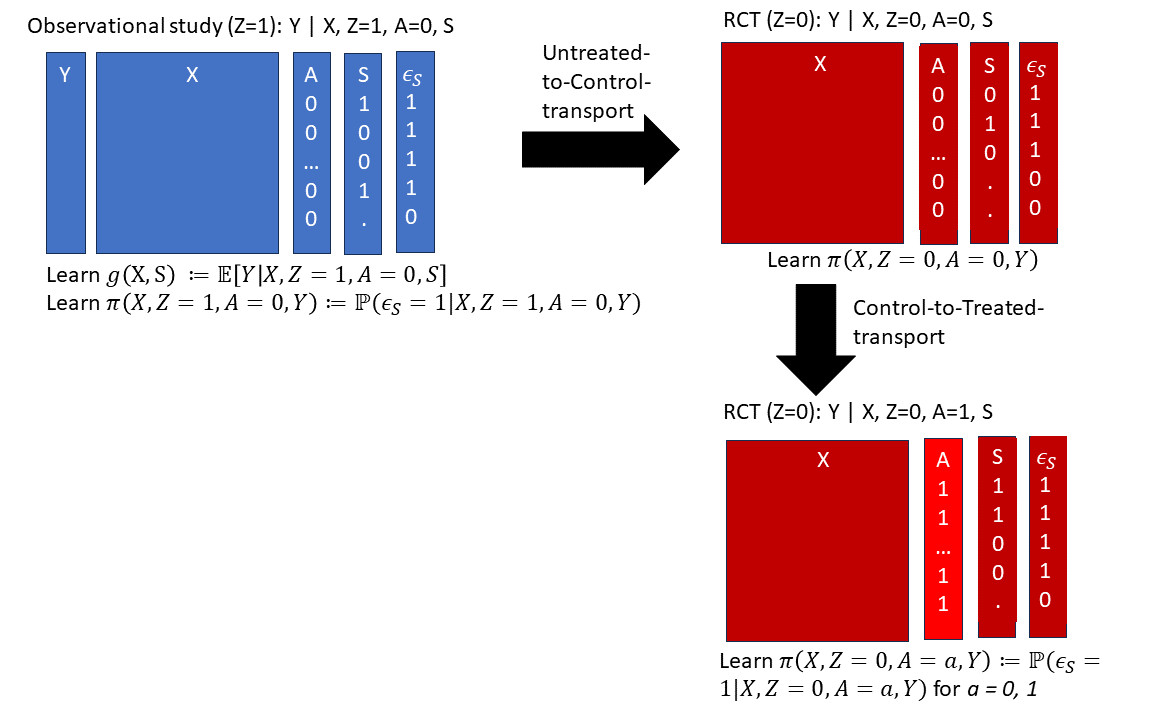}
    \caption{Transportability: Learning about target outcomes $Y$ for treatment $A=1$ in a phase 3 target trial $Z=0$ using information from untreated $A=0$ participants in an observational study $Z=1$, where the $S$ sampling probability $\pi(X,Z=1,A=0,Y) := P(\epsilon_S = 1 | X,Z=1,A=0,Y)$, and, in the presence of right-censoring of $T$, $\pi(X,Z=1,A=0,Y)$ is replaced with $\pi(X,Z=1,A=0,\tilde T,\Delta) := P(\epsilon_S = 1 | X,Z=1,A=0,\tilde T, \Delta)$.  While the $S$ sampling probability may be approximately known by investigator control of the sampling design, it is still learned/estimated to improve efficiency and to accommodate happenstance missing data.
    A bias function is used in each step of the transport [see definitions (\ref{eq: 32}) and (\ref{eq: 33})] to specify uncertainty in transportability.
    }
    \label{fig:transportschema}
\end{figure}

In this step, the researcher specifies available knowledge about the candidate surrogate $S$ as a valid surrogate for the target outcome of interest as pertinent to the specific treatment under development vs. the control arm.
This knowledge will be needed for devising an approach to transporting knowledge learned about the relationship of $(X,Z=0,A=0,S)$ with $Y$ in the observational study (in a causal sense considered in two parts) to two new settings lacking direct empirical data. Figure \ref{fig:transportschema} diagrams the two parts. Part 1, ``Untreated-to-Control-transport  across studies," transports 
the relationship of $(X,S)$ with $Y$ from the (all) untreated observational study population 
($A=0,Z=1$) to the untreated/placebo arm of the phase 3 target trial ($A=0,Z=0$).
Part 2, ``Control-to-Treated-transport in the phase 3 trial," transports 
the relationship of $(X,S)$ with $Y$ from the 
untreated/placebo arm of the phase 3 target trial
($A=0,Z=0$) to the treated arm of the phase 3 target trial
($A=1,Z=0$).

Untreated-to-Control-transport  addresses the need to bridge knowledge of outcomes learned for the untreated in the observational study to the untreated in the phase 3 study, accounting for potentially different distributions of baseline variables and of the candidate surrogate.  Control-to-Treated-transport addresses the need to take the bridged knowledge of outcomes for the untreated in the phase 3 study to the treated in the phase 3 study, addressing the issue that the candidate surrogate may relate to the target outcome differently in the treated and the untreated.  Conceptualizing the bridging in two distinct parts has advantage of aiding examination of assumptions and designing interpretable sensitivity analyses.  In the Discussion we note an alternative single-part approach.

\begin{figure}[h!]
    \centering
    \includegraphics[width=\textwidth]{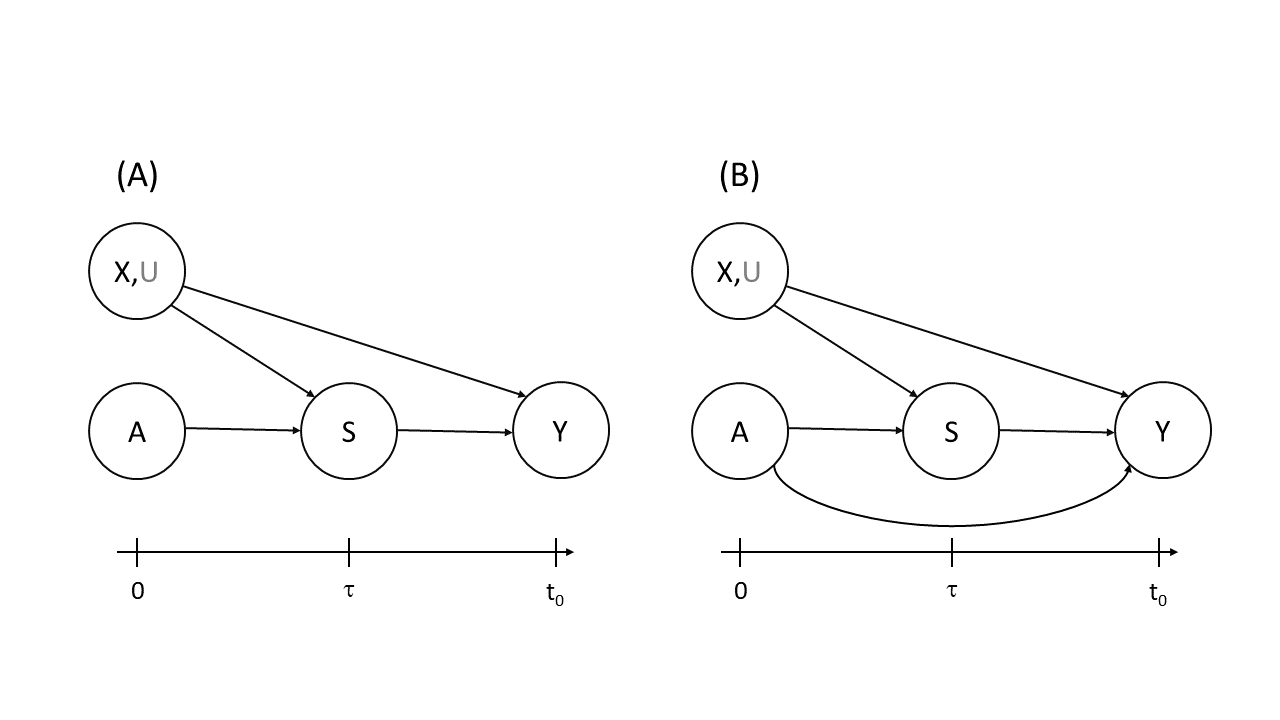}
    \caption{Causal models for the relationship of treatment $A$, baseline covariates $X$, surrogate $S$, and target outcome $Y$ in the phase 3 target trial $Z=0$: (A) Model expressing a perfect surrogate ($u^{CT}(X,S)=0$), (B) Model expressing an imperfect surrogate via a non-zero bias function $u^{CT}(X,S)$ where $A$ affects $Y$ through mechanisms not captured by $S$. $U$ indicates potential unmeasured confounders of the effect of $S$ on $Y$.}
    \label{fig:dag_phase3}
\end{figure}

Typical pre-requisite knowledge for a candidate surrogate endpoint to hold promise for being valid and able to accomplish the objectives include: (1) the endpoint is measured accurately and precisely, with low measurement error; (2) the endpoint has broad inter-individual variability across treated and untreated persons (for the populations studied); (3) the endpoint is strongly associated with the target outcome in natural history contexts including the observational study $Z=1$ within levels of baseline covariates $X$; and (4) the endpoint is connected to putative causal pathway mechanisms of effectively treating the disease.  
\comment{Lei: Do we want to say the endpoint captures the effect on the causal pathway? PG: Added the term causal pathway.}
Evidence for (4) can be most compelling when generated from studies that directly manipulate/assign the surrogate endpoint such as experiments that can be conducted in animal models.
Figure \ref{fig:dag_phase3} shows two causal models for the surrogate endpoint in the randomized $Z=0$ study.  Panel (A) expresses a perfect surrogate causal model, as defined by equal conditional means under either
treatment assignment $A=1$ vs. $A=0$ 
[i.e., $E[Y(1)|X=x,Z=0,S(1)=s] = E[Y(0)|X=x,Z=0,S(0)=s]$], expressed in detail as assumption A6 below with bias function $u^{CT}(X,S)=0$.  Panel (B) expresses an imperfect surrogate causal model, where the treatment $A$ has an additional effect on $Y$ not mediated through $S$, a situation where A6 is needed with non-zero bias function to account for the incomplete mediation.  

Measuring the same $X$ and $S$ at the same time points in the same way in the $Z=1$ and $Z=0$ studies is basic knowledge informing Untreated-to-Control-transport, as is designing the phase 3 study to take place in a similar population as the observational study. For informing Control-to-Treated-transport, if phase 3 target-outcome trials were available, then a large literature of statistical methods for surrogate endpoint evaluation could be applied [e.g., \citep{Buyseetal2016,xie2019systematic,weir2022informed,wheaton2023using}], where, for example, assessing the nearness of controlled direct effects (at $S=s$) of $A$ on $Y$ (which compare risk under assignment to $A=1$ vs. $A=0$ holding $S=s$ fixed for both $A=1$ and $A=0$) to zero would be valuable knowledge for specifying the causal model \citep{JoffeGreene2009,GilbertFongKennyCarone2022}.  However, for our set-up with no previous phase 3 trials, researchers will need to leverage domain knowledge of the specific disease, the specific intermediate outcomes $S$ and treatment, and the specific target outcome.

\subsection{Step 2: Define the causal parameter of interest}
\label{Step2}

With $S(a)$ and $Y(a)$ potential outcomes under randomization assignment $a=0,1$ in the phase 3 trial $Z=0$, the causal parameter of interest is a contrast of $E[Y(1)|Z=0]$ and $E[Y(0)|Z=0]$, such as the average treatment effect $E[Y(1)|Z=0] - E[Y(0)|Z=0]$. 
\comment{ \id{What is the reason to choose the trial population as the target of inference? Depending on the source of the data for the observational study, someone might argue that the obs study is a more policy-relevant population (e.g., if the obs study is from electronic health records).} 
{\color{purple} PG 1-26-24: I agree that for some applications it would be of interest to make inference for the observational study population.  However for the particular objective addressed in this manuscript -- provisional approval via a surrogate-endpoint comparative RCT -- I think we can say why the RCT is the target of inference.  I have added text endeavoring to explain this. Optionally, we could have a brief supp section describing how a very similar approach (with one of the bias functions altered) could be used for the observational study population as the target for inference.} }
To fit the GBS vaccine case study, throughout we consider the multiplicative vaccine efficacy or treatment efficacy (TE) contrast TE = $1 - 
E[Y(1)|Z=0] \slash E[Y(0)|Z=0].$  The development considers estimation and inference for each of $E[Y(1)|Z=0]$ and $E[Y(0)|Z=0]$, implying it applies for any preferred contrast of $E[Y(1)|Z=0]$ and $E[Y(0)|Z=0]$.

We focus on the phase 3 trial population as the target for inference  because for our provisional approval objective the normative approach in practice defines the success criterion (indicating qualification for provisional approval) in terms of a causal treatment effect on the surrogate endpoint in the phase 3 trial.  In vaccine studies this approach, frequently used by the U.S. FDA, is often referred to as clinical immunobridging (e.g., \cite{krause2022making}).  The randomization in the phase 3 trial not only facilitates achieving valid estimation/inference for an average treatment effect causal parameter, but also is a key factor for attaining widespread credibility of the results that is important for uptake of the provisionally approved treatment.  Alternatively, the observational study population could be the target of inference (studying a contrast of $E[Y(1)|Z=1]$ vs. $E[Y(0)|Z=1]$), which will frequently be of interest in its own right. The methods in this manuscript can be readily adapted to make inference for the observational study population.

\comment {The causal estimand here is well defined and well thought, which intends to predict treatment effect based on 1) information on surrogate and endpoint and target outcome from observational studies, and information on the surrogate endpoint from the randomized study(ies). 

Are we interested in an alternative goal to establish/validate the surrogacy of the endpoint $S$ (as a surrogate endpoint of target outcomes, often independent of treatments)? If we are, we may define the average causal medication effect across different treatments.  

PG: This strikes me as a new objective, where it would be challenging to add this to the scope of this manuscript, so one thought is to add `alternative objectives'/parameters to the Discussion.
}

\subsection{Steps 3 and 4: Translate the causal parameter into a statistical estimand; identify conditions under which the statistical estimand equals or bounds the causal parameter } 
\label{RoadmapStep3}

We combine Steps 3 and 4 into one section given their tight inter-dependency.  After listing identifiability conditions (Step 4), we describe how under these assumptions Step 3 attains.  The conditions use two known user-specified bias functions for sensitivity analysis:
\begin{eqnarray*}
u^{UC}(x,s) & := & E[Y(0)|X=x,Z=1,S(0)=s] - E[Y(0)|X=x,Z=0,S(0)=s],\\
u^{CT}(x,s) &:= & E[Y(1)|X=x,Z=0,S(1)=s] - E[Y(0)|X=x,Z=0,S(0)=s],
\end{eqnarray*} 

\noindent with UC standing for Untreated-to-Control-transport and
CT standing for Control-to-Treated-transport.  
The two bias functions can both be written as statistical estimands:
\begin{eqnarray}
u^{UC}(x,s) & = & E[Y|X=x,Z=1,A=0,S=s] - E[Y|X=x,Z=0,A=0,S=s], \label{eq: 32} \\
u^{CT}(x,s) & = & E[Y|X=x,Z=0,A=1,S=s] - E[Y|X=x,Z=0,A=0,S=s] \label{eq: 33}
\end{eqnarray} 

\noindent where (\ref{eq: 32}) holds by causal consistency A1 and because $A=0$ for all $Z=1$ participants, and (\ref{eq: 33}) holds by A1 and because the $Z=0$ study is randomized (A2).

The UC bias function is not identified from the observed data, because $Y|X,Z=0,A=0,S$ is not directly studied in the $Z=0$ study, and is thus treated as a fixed function expressing a degree of residual unaccounted for differences in background risk after accounting for differences in $(X,S)$.  The special case $u^{UC}(x,s)=0$ for all $(x,s)$, indicating no residual bias, is not of central interest for our provisional approval application where lower-bound estimation of TE is a central concern.

\comment{
{\color{orange} DF 2-18-24: But if CT is identified 
in the phase 3 trial why isn't UC?  Would be underpowered
as you point out.

CT bias function looks like a direct effect.
For vaccines we think of direct effects as being B/T cells
and indirect effect via antibody.  But here only antibodies from vaccinted mom (or unvaccinated mom in Z=1) get to fetus so no B/T cell  direct effect of vaccination. }

{\color{purple} PG: 3-12-24: good point that for the GBS example the fact that $S$ is passively transferred antibody is a rationale for positing that the CT bias function shouldn't be super far away from zero.  Although something that needs to be done is check that kinetics of antibody are similar for vaccine-induced vs. infection-induced.}

In Z=1 there is a relationship between X,S, one could see if
the same relationship held in Z=0.  This should have more power and 
if different relationships, might want to increase
the bias function $u^{UC}$ thinking if $f(X,S|Z=0) \neq f(X,S|Z=1)$
then likely $f(Y|X,Z=0,S) \neq f(Y|X,Z=1,S)$
{\color{purple} PG: 3-12-24: OK, we can put this thought in the section that discusses how to set the UC bias function for the GBS example.  I put a bit of material there from your comment here. }

}

While the CT bias function would be identified from the observed data if the phase 3 study collects data on $Y$, because even with such data collection the study would be vastly under-powered to estimate $u^{CT}(x,s) $, it is also treated as a fixed user-specified function expressing the quality of the candidate surrogate $S$, with the ideal/perfect surrogate attaining $u^{CT}(x,s)=0$.  Again this special case is not of central interest given that in our provisional approval application the inference on VE needs to be robust to a less-than-perfect surrogate $S$.

We list the identifiability conditions (Step 4), describe how they link the causal parameters to statistical estimands (Step 3), and then discuss the identifiability conditions.  

\subsubsection{Step 4: Identifiability assumptions}

\comment{
I suggest moving section "3.4  Identifiability assumptions" to before '3.3 Steps 3 and 4' to enhance the readability. }

\begin{itemize}
\item[A1] {\it Causal consistency:} For each individual $i$ in the phase 3 study $Z=0$, the observable surrogate and observable target outcome under treatment $A_i=a$ equals that individual's counterfactual surrogate and counterfactual target outcome under the same treatment, i.e., $A_i=a$ implies $S_i=S_i(a)$ 
 and $Y_i=Y_i(a)$ for $a=0,1$.  For each individual $i$ in the observational study $Z=1$, the same results attain for $a=0$.

\item[A2] {\it Conditional mean exchangeability in the phase 3 trial (over $A$):} Among randomized individuals (in study $Z=0$), the potential target outcome mean under treatment $a$ is independent of treatment,
conditional on baseline covariates $X$, i.e., 
$E[Y(a)|X=x,Z=0,A=a] = E[Y(a)|X=x,Z=0]$ for each $a=0,1$ and every
$x$ with positive density $f_{(X,Z)}(x,z=0) > 0$.

\item[A3] {\it Positivity of treatment assignment in the phase 3 study:} In the phase 3 study, the probability of being assigned each treatment $a=0,1$, conditional 
on the covariates $(X,S)$ used in assumption A2, is positive, i.e., $1 > P(A=a|X=x,Z=0,S=s) > 0$ for each $a=0,1$ and every $(x,s)$ with positive joint density $f_{(X,Z,S)}(x,z=0,s) > 0$.

\item[A4] {\it {\bf Comparability Assumption} -- Conditional mean exchangeability of the untreated with bias function from the observational study to the phase 3 study (exchangeability over $Z$ in the untreated $A=0$):} 
$E[Y(0)|X=x,Z=1,S(0)=s] = E[Y(0)|X=x,Z=0,S(0)=s] - u^{UC}(x,s)$ for every $(x,z=0,a=0,s)$ with positive joint density $f_{(X,Z,A=0,S)}(x,z=0,a=0,s) > 0$.

\item[A5] {\it Positivity of being enrolled in the observational study:} The probability of enrolling in the observational study, conditional  on the covariates $(X,S)$ used in assumptions A2, A3, and A4, is positive, i.e., $1 > P(Z=1|X=x,S=s) > 0$ for every $(x,s)$ with 
positive joint density $f_{(X,S)}(x,s) > 0$.

\item[A6] {\it {\bf Surrogate Assumption}-- Correct specification of the CT bias function that provides exchangeability over $A$ in the phase 3 study:} 
$E[Y(1)|X=x,Z=0,S(1)=s] = E[Y(0)|X=x,Z=0,S(0)=s] + u^{CT}(x,s)$ for every $(x,z=0,a=0,s)$ with positive joint density $f_{(X,Z=0,A=0,S)}(x,z=0,a=0,s) > 0$.


\item[A7] {\it Missing at random candidate surrogate in both studies:} The probability of observing $S$ does not depend on the value of $S$, i.e., with
\begin{eqnarray}
    \pi(X,Z,A,\tilde T,\Delta) := P(\epsilon_S = 1 | X,Z,A,\tilde T,\Delta), \label{eq: pidef}
\end{eqnarray}
$\pi(X,Z=1,A=0,S,\tilde T,\Delta) = 
\pi(X,Z=1,A=0,\tilde T,\Delta)$ and 
$\pi(X,Z=0,A,S,\tilde T,\Delta) = 
\pi(X,Z=0,A,\tilde T,\Delta)$.  Note that because target outcome data $(\tilde T,\Delta)$ are not available/used for the $Z=0$ study participants, $\pi(X,Z=0,A,\tilde T,\Delta)$ actually equals $P(\epsilon_S = 1 | X,Z=0,A)$; henceforth we leave this tacit in the notation.

\item[A8] {\it Random right-censoring of the target outcome in the observational study:} In the observational study, no participants are right-censored through the final time point $t_0$ or right-censoring is random within levels of baseline covariates, i.e., 
$T \perp C | X, Z=1, A=0$.  



\end{itemize}

\comment{
{\color{orange} DF 2-18-24:  
Might the vaccine induce $S$ larger than seen in the observational
study?  So you might want an assumption of common support, though
this might be violated.   But even if violated one could
make a conservative extrapolation assumption where if sm is the largest S seen in the observational study one assumes  $P(Y=1|S>sm) = P(Y=1|S=sm)$ and that it applies to the vaccinees in the trial.}

{\color{purple} PG 3-12-24: A3 is the common support assumption, which definitely could be violated. Agree the conservative extrapolation assumption makes sense, which is equivalent to assigning a maximum tied value of $S$ in the common support and chugging through assuming A3, where the argument for being conservative is compelling.  Added a comment on this in the Discussion.}
}

\subsubsection{Step 3: Connecting the causal parameters to the statistical estimands}
Define $g(X,S) := E[Y|X,Z=1,A=0,S]$.
For the Untreated-to-Control-transport, the following equation holds under A1, A4, A5, A8:
\begin{eqnarray}
E[Y(0)|Z = 0] & = & E\big\{E[ g(X,S) | X, Z = 0, A=0] \mid Z=0\big\} \nonumber \\
& & - E\big\{E[\mu^{UC}(X,S) | X, Z=0, A=0]\mid Z=0\big\},  
\label{eq: Untreatedtransport}
\end{eqnarray}

\noindent which is identified from the data for each user-specified $u^{UC}(X,S)$ function.


The last step establishing identifiability addresses the fact that $S$ is only measured in a random sample.  From \citet{RosevanderLaan2011},
$g(X,S)$ is identified as:
\begin{eqnarray}
    g=\arg\min_{\tilde g}E\left[\frac{\epsilon_S}{P(\epsilon_S=1 \mid X,A=0,\tilde T, \Delta)}L(\tilde g)(X,S,\tilde T, \Delta)\,\bigg|\, A=0\right], \label{eq: rosevan}
\end{eqnarray}
where $L(\tilde g)(X,S,\tilde T, \Delta)$ is the loss function as if $S$ had been measured for all participants. 

For the Control-to-Treated-transport, the following equation holds under A1--A6, A8:
\comment{
\begin{eqnarray}
E[Y(1)|Z=0] = E[g(X,S) |Z=0,A=1] + E[u^{CT}(X,S) - u^{UC}(X,S)|Z=0,A=1]. \label{eq: Controltotreatedtransport}
\end{eqnarray}
}
\begin{eqnarray}
E[Y(1)|Z = 0] & = &E\big\{E[ g(X,S) | X, Z = 0, A=1] \mid Z=0\big\} \nonumber \\
& & + E\big\{E[\mu^{CT}(X,S) - \mu^{UC}(X,S)| X,  Z=0,A=1] \mid Z=0\big\}
\label{eq: Controltotreatedtransport}
\end{eqnarray} 

\noindent which is identified from the data for each pair of user-specified functions $u^{UC}(X,S)$ and $u^{CT}(X,S)$.  Then equation (\ref{eq: rosevan}) establishes identifiability of $E[Y(1)|Z=0]$.
%
\comment{
\noindent Again the missing at random assumption A7 completes
a linkage of the causal parameter $E[Y(1)|Z=0]$ to a statistical estimand:
\begin{eqnarray}
E[Y(1)|Z=0] & = & E[g(X,Z=1,S,\epsilon_S=1)|Z=0,A=1,\epsilon_S=1] \nonumber \\
& & + E[u^{CT}(X,S) - u^{UC}(X,S)|Z=0,A=1,\epsilon_S=1]. \label{eq: ControltotreatedtransportMAR}
\end{eqnarray}
%
}

The two equations (\ref{eq: Untreatedtransport}) and (\ref{eq: Controltotreatedtransport}) are designed such that
$g(X,S)$, which conditions on being in the observational study $Z=1$ and is
the only direct source for learning about the surrogate-target outcome relationship, 
is the key regression for estimating  
treatment efficacy in the new setting $Z=0$.  The formulas reveal that, in the special case with both bias functions equaling zero, high treatment efficacy TE is attained if both (1) 
$g(X,S)$
is monotone non-increasing in $S$ and decreases by a large amount; and (2) the distribution of the surrogate $S$ in the $Z=0$ study is stochastically much higher in the treated than untreated control $A=1$ vs. $A=0$.  Vice versa, either
$g(X,S)$ independent of $S$ (i.e., $g(X,S)=g(X)$ such that $S$ is not a correlate of risk in the $Z=1$ study), or $S|X,Z=0,A=1 =^d S|X,Z=0,A=0$ (no treatment effects on the surrogate in the $Z=0$ study), imply TE = 0. Appendix A shows the derivation of 
(\ref{eq: Untreatedtransport}) and (\ref{eq: Controltotreatedtransport}) detailing which assumptions are needed for each step.



A1--A5 are similar to identifiability conditions (I)--(V) respectively of \cite{dahabreh2023sensitivity}, except in our problem the target setting for transporting the treatment effect ($Z=0$) is the randomized trial, not the reverse as in \cite{dahabreh2023sensitivity}, and A3, A4, A5 involve intermediate outcomes $S$ that \citet{dahabreh2023sensitivity} did not consider.  
Also, in A4 we include the bias function $u^{UC}(X,S)$ in the assumption, where
setting $u^{UC}(X,S) = 0$ yields the standard conditional mean exchangeability assumption across studies for the untreated \citep{rudolph2017robust,dahabreh2023sensitivity}.  
A5 is an overlap condition making possible borrowing knowledge on background risk in the observational study to apply to the untreated control arm in the phase 3 study.

A6 is new for our problem, which exchanges/transports knowledge of the relationship of the surrogate and target outcome from the untreated control arm to the treated.
Setting $u^{CT}(X,S) = 0$ defines the ``complete mediation" condition that is a causal version of 
Prentice's (1989)\nocite{Prentice1989} third criterion for a valid surrogate endpoint that was also considered in \cite{PriceGilbertvanderLaan2018} (Theorem 3).  
Specifying non-zero $u^{CT}(X,S)$ provides a sensitivity analysis acknowledging imperfection of the surrogate.
Use of A6 relies on A3 which requires that the distribution of the surrogate endpoint has overlapping support across the treated and control arms of the phase 3 study.  A7 and A8 address missing data on $S$ and missing data on $Y$, respectively. 

For our purpose of informing provisional approval, it is of interest to specify both $u^{UC}(X,S)$
and $u^{CT}(X,S)$ to make it more difficult to meet the approval success criteria, as a means to inculcate conservatism given A4 and A6 are untestable and hence being conservative is valuable for lowering the risk of provisionally approving a treatment that turns out to have poor performance against the target outcome.  Section \ref{RoadmapStep6} discusses specifications of the bias functions.

\subsection{Step 5: Estimate the statistical estimand}
\label{roadmapstep5}

Estimation requires a technique for accommodating missing data for $S$ in both studies, as well as for right-censoring of $T$ before $t_0$ in the $Z=1$ study.  
 We address missing values of $S$ with inverse probability sampling (IPS) weighted complete-case estimation, with $\pi(X,Z,A,\tilde T,\Delta)$ as defined at (\ref{eq: pidef}).
 Because in both studies $Z=0,1$ the investigator designs a plan for sampling the set of participants from whom to measure $S$, with the sampling design depending on
$(X,Z=1,A=0,\tilde T,\Delta)$ for the observational study and on $(X,Z=0,A)$ for the randomized study,
it is generally attainable to correctly model $\pi(\cdot)$. 
The statistical estimands [RHSs of equations (\ref{eq: Untreatedtransport}) and (\ref{eq: Controltotreatedtransport})] that link to the causal target parameters $E[Y(0)|Z=0]$ and $E[Y(1)|Z=0]$ under the causal assumptions can be written as
\begin{eqnarray} 
\theta_a := E\big\{E[ g^*_a(X,S) | X, Z = 0, A=a] \mid Z=0\big\}, a=0,1, \label{eq: targetparameter}
\end{eqnarray}

\noindent where 
\comment{
$g^*_a(x, s) := E[ Y^*|X=x,Z=1,A=0,S=s]$ with
$Y^* = Y - \mu^{UC}(X,S)$ for $a=0$ recovering the identification formula for $E[Y(0)\mid Z=0]$ and $ Y^* = Y +\mu^{CT}(X,S) - \mu^{UC}(X,S)$ for $a=1$ recovering the identification formula for $E[Y(1)\mid Z=0]$.
}
$$g^*_0(x, s) := E[Y - \mu^{UC}(X,S) |X=x,Z=1,A=0,S=s]$$

\noindent recovers the identification formula for $E[Y(0)\mid Z=0]$ and 
$$g^*_1(x, s) := E[Y +\mu^{CT}(X,S) - \mu^{UC}(X,S) |X=x,Z=1,A=0,S=s]$$ 
\noindent recovers the identification formula for $E[Y(1)\mid Z=0]$.

\subsubsection{Plug-in Estimator}

Define the plug-in estimator for each $a=0,1$ as 
    \begin{eqnarray}
        \widehat\theta_{a,\mbox{\text{\tiny plug-in}}}=
        \frac{1}{n_{RCT}}\sum_{i=1}^n (1 - Z_i) \widehat E[ {\widehat g^*_a(X,S)} | X_i, Z_i = 0, A_i=a], \label{eq: plug-in}  
    \end{eqnarray}
where $\widehat g^*_a(X,S)$ is obtained using the loss function in equation (\ref{eq: rosevan}), which can be implemented with any regression estimator using weights $1\slash \widehat \pi(X_i,Z_i,A_i,\tilde T_i,\Delta_i)$. 
If the bias functions $u^{UC}(X,S)$ and $u^{CT}(X,S)$ are set to known constants
$u^{UC}$ and $u^{CT}$, then $\widehat g^*_a(X,S)$ is obtained as
$\widehat g(X,S) - u^{UC}$ and $\widehat g(X,S) + u^{CT} - u^{UC}$ for $a=0$ and $a=1$, respectively, based on a single regression estimator  $\widehat g(X,S)$.
 The outer expectation estimate $\widehat E[\cdot]$ can be obtained for $a=0$ by regressing $\widehat g^*_0(X,S)$ on $X_i$ including individuals with $Z_i=0$ and $A_i=0$, and for $a=1$ by regressing 
 $\widehat g^*_1(X,S)$ on $X_i$ including individuals with $Z_i=0$ and $A_i=1$, where both of these regressions only include participants with $\epsilon_{Si}=1$ and include the IPS weights $1\slash \widehat \pi(X_i,Z_i,A_i,\tilde T_i,\Delta_i)$. 

After Section \ref{NPEIF} we restrict to constant bias functions, which
simplifies the discussion given that $g(X,S)$ is the only outcome regression needing estimation, and it provides the most easily interpreted sensitivity analysis.

\subsubsection{Nonparametric efficient estimator}
\label{NPEIF}

To develop an efficient estimator, we follow a general approach outlined by \cite{RosevanderLaan2011}.  First, assume a hypothetical data structure with $\epsilon_S=1$ with probability one. The efficient influence function (EIF) then would be $\varphi_a(O;\eta) = $
\begin{eqnarray}
    & &\frac{I(Z=1)}{P(Z=0)}\frac{I(A=0)}{P(A=a\mid X, Z=0)}\frac{P(Z=0, A=a\mid X,S)}{P(Z=1, A=0\mid X,S)}\{ Y + a\mu^{CT}(X,S) - \mu^{UC}(X,S) -  g^*_a(X, S)\} \nonumber \\
    & &+\frac{I(Z=0)}{P(Z=0)}\frac{I(A=a)}{P(A=a\mid X, Z=0)}\{ g^*_a(X, S) - E[g^*_a(X, S)\mid X, Z=0, A=a]\} \nonumber \\
    & &+\frac{I(Z=0)}{P(Z=0)}\{E[ g^*_a(X, S)\mid X, Z=0, A=a] - \theta_a\}, \label{eq: EIFcomplete}
\end{eqnarray}
where $\eta$ is used to denote the nuisance parameters appearing in the EIF. Results in \cite{RosevanderLaan2011} show that, when $P(\epsilon_S=1) < 1$, the EIF can be constructed as:
\begin{eqnarray}
    \phi_a(O;\eta) & = & \frac{\epsilon_S}{P(\epsilon_S=1 \mid X,Z,A,\tilde T, \Delta)}\varphi(O;\eta)\\ 
    & & + \left\{1-\frac{\epsilon_S}{P(\epsilon_S=1 \mid X,Z,A,\tilde T, \Delta)}\right\}E[\varphi(O;\eta)\mid \epsilon_S=1, X,Z,A,\tilde T,\Delta]. \label{eq: EIF}
\end{eqnarray}


These calculations motivate the construction of a one-step estimator through the following steps:
\begin{enumerate}
    \item Construct estimators of all  nuisance parameters using regression. For each $g^*_a$, 
    use the loss function in (\ref{eq: rosevan}). The IPS weight function  
    $1\slash \pi(X, Z, A, \tilde T,\Delta)$ may be estimated by logistic regression. For the other nuisance parameters, one choice uses the empirical estimator of $P(Z=0)$,
    parametric or superlearner regression for each
    $P(A=a \mid X, Z=0)$,
    and IPS-weighted parametric or superlearner regression for each $P(Z=0, A=a\mid X,S)$ and $P(Z=1, A=0\mid X,S)$.



    
    Let the nuisance estimates be denoted $\hat\eta$.

\item Define the plug-in estimator 
    $\hat\theta_{a,\mbox{\text{\tiny plug-in}}}$ as in equation (\ref{eq: plug-in}).
    
    \item For observations with $\epsilon_{S,i}=1$, compute $\varphi_a(O_i;\hat\eta)$. Among these observations, regress $\varphi_a(O;\hat\eta)$ on $(X,Z,A,\tilde T, \Delta)$ using the plug-in estimate of each $\theta_a$. Compute the predictions from this regression for all observations, and compute $\phi_a(O_i,\hat\eta)$ for all observations. 

    \item Define the one-step estimator as 
\begin{eqnarray}
    \hat\theta_{a,\mbox{\text{\tiny one-step}}}=\hat\theta_{a,\mbox{\text{\tiny plug-in}}} + \frac{1}{n}\sum_{i=1}^n \phi_a(O_i,\hat\eta).
    \label{eq: one-step}
    \end{eqnarray}
\end{enumerate}

\noindent For the plug-in estimator to be consistent, in the observational study each regression estimator $\widehat g^*_a(X,S)$ must be consistent for $g^*_a(X,S)$, for $a=0,1$, and 
$\hat{\pi}(X,Z,A,\tilde T,\Delta)$ must be consistent for $\pi(X,Z,A,\tilde T,\Delta)$ in both studies $Z=0,1$.
For the efficient estimator to be consistent, the additional nuisance estimates must also be consistent for their respective parameters. The efficient estimator can gain efficiency by accounting for information in the phase-one data $(X,Z,\tilde T,\Delta)$ that is not included in the plug-in estimator.
Under assumptions A1--A8, consistent estimation of $\pi(\cdot)$, and by selecting nuisance regression estimators that are not too data-adaptive such that
all of the nuisance parameter estimators meet convergence rate conditions, then
both the plug-in and one-step efficient estimator have asymptotically normal distributions. 


Computation of the above estimator requires estimating $E[g^*_a(X, S)\mid X, Z=0, A=a]$, which is in principle not identified since it relies on observing $S$ for everyone in the sample. However, since the sampling mechanism $\pi(X,Z,A,\tilde T, \Delta)$ is known, this can be overcome by using weighted regression as in (\ref{eq: rosevan}).

Both estimators have the limitation of being designed to work best for an unbounded $Y$ (i.e. a continuous target outcome), as they do not take into account the structural knowledge that $Y \in \lbrace 0, 1 \rbrace$.  A targeted minimum loss-based estimator (e.g., 
\citet{BenkeserCaroneGilbert2018}) could possibly provide improved finite-sample performance over the one-step estimator because it could be designed to enforce the structural knowledge.

\comment{
The estimators can be augmented to gain back this efficiency, for example by modifying (\ref{eq: ORest0}) to 
\begin{eqnarray}
& & \hspace*{-.4in} \widehat E[Y(0)|Z=0] = \left\{ \sum_{i=1}^{n}  
 \frac{\epsilon_{Si}(1-Z_i)(1-A_i)}{\widehat \pi(X_i,Z_i,A_i,Y_i)} \right\}^{-1} 
\sum_{i=1}^{n}  (1-Z_i)(1-A_i) \nonumber \\
& & \hspace*{-.4in} \left\{ \frac{\epsilon_{Si}}{\widehat \pi(X_i,Z_i,A_i,Y_i)} \left\{ \widehat g(X_i,S_i) - u^{UC}(X_i,S_i) \right\}  
- \left( \frac{\epsilon_{Si}}{\widehat \pi(X_i,Z_i,A_i,Y_i)} - 1  \right) \widehat h_1(X_i,Z_i=1,S_i) \right\}  \label{eq: ORestaug}
\nonumber \\
\end{eqnarray}

\noindent where $\widehat h_1(X,Z=1,S)$ is an estimator of
$h_1(X,Z=1,S) := E[g(X,S) - u^{UC}(X,S) | X,Z=0,A=0,\epsilon_S=1]$.
The augmented estimator for $E[Y(1)|Z=0]$ is the same with $1-A_i$ replaced with $A_i$ and 
$u^{UC}(X,S)$ replaced with
$-u^{CT}(X,S) + u^{UC}(X,S)$.
}

\subsubsection{Estimation of the outcome regression: Engine of the transport}
\label{engine}

We now discuss the key issue of how to estimate the outcome regression
$g(X,S)$ in the $Z=1$ study.  Without right-censoring of $Y$, a parametric or semiparametric model for a dichotomous outcome completely observed could be used (e.g., logistic regression or partially linear logit model).  With right-censoring of $Y$,
a parametric or semiparametric survival model that readily 
yields an estimator of $g(X,S)$ could be applied.
However, because consistent estimation of TE depends on a correctly specified model for $g(X,S)$, it is desirable to seek flexible estimation 
of $g(X,S)$.  Ensemble-based super-learning \citep{vanderLaanetal2007} provides one approach, as considered by \citet{PriceGilbertvanderLaan2018} 
for obtaining the estimated optimal surrogate (EOS) that is defined as the optimal estimator of $g(X,S)$ by minimizing cross-validated risk.
\comment{ In their set-up, the data in the $Z=1$ study are available for both the treated and control arms $A=1$ and $A=0$, such that an EOS was developed encompassing both $g_{a=0}(X,S)$ and $g_{a=1}(X,S)$, whereas for our context only $g(X,S)$ is relevant/exists.  
\citet{PriceGilbertvanderLaan2018} generated the EOS in two steps, first computing the superlearner estimator of ($g(X,Z=0,S), g(X,S)$), and secondly bias-correcting this estimator by solving an efficient influence score equation for estimation of the average treatment effect on the target outcome in the original study.  However, as noted, because treated participants are not included in the original study, the second update step does not apply in our setting.
}

The theoretical results of \citet{PriceGilbertvanderLaan2018} did not consider missing data in $S$; for our problem this is needed. One approach employs IPS weighted superlearner \citep{RosevanderLaan2011}, including the IPS weights in all of the individual learners that are members of the selected superlearner library.  In addition, these weights are included in the estimation of performance metrics quantifying classification accuracy of the EOS, such as the cross-validated area under the ROC curve.
If $T$ is subject to right-censoring, then the weights can also incorporate estimates for each observed failure event the reciprocal probability of not being right-censored by their failure time \citep{robins2000correcting}. 
Another possible estimator in the right-censoring case is a debiased superlearner estimator 
\comment{Can the sas procedure PROC CAUSALTRT be used to fit the models here? PG: I don't have experience with this sas procedure.}
\citep{wolock2022framework}, implemented in the R package {\it survML} available at Charles Wolock's Github page. 
Given the target outcome is a rare event, estimators constrained by a maximum possible event probability would be expected to provide finite-sample precision gains \citep{BalzervanderLaan2013,BenkeserCaroneGilbert2018}.

\comment{
Put in Nima approach... 

This approach does respect the bounds of $Y = I(T \le t_0) \in \lbrace 0, 1\rbrace$ but it does not yet accommodate right-censoring of $T$.

More nonparametric   HAL    superlearning + cross-fitting ...  from Nima ms

}

\subsubsection{Estimation under different statistical models}

Two examples of statistical modeling approaches are as follows.
In Statistical Model 1, achieving a level of the candidate surrogate above a fixed threshold is important for achieving treatment benefit against the target outcome, whereas in Statistical Model 2, the whole spectrum of levels of the candidate surrogate impact the level of treatment benefit.  
This choice leads to different estimators of the statistical estimands in Step 5, and are considered for the GBS vaccine case study.

Under Statistical Model 1, the regression model or ensemble of regression models can be selected to emphasize models that find cut-points of $S$ predictive of especially low risk. Learners that consider many possible cut-points that divide $S$ into 2- or 3-level discrete categorical variables may be of interest [e.g., \cite{van2023nonparametric}], as are tree-based learners and parametric threshold or hinge models \citep{fong2017chngpt}.
Under Statistical Model 2, one approach calculates the EOS $\widehat g(X,S)$ with a regression model or ensemble of regression models selected to emphasize relationships between $(X,S)$ and $Y$ that can depend on the whole distribution of $S$ (e.g., generalized additive models, multivariate adaptive regression splines).



\subsubsection{Intercurrent events}
\label{ICEsec}

Intercurrent events (ICEs) in our context are events that occur after enrollment/treatment initiation and affect either the interpretation or existence of the surrogate endpoint and/or target outcome \citep{saddiki2020primer}. Mapping of the causal parameters to statistical estimands requires dealing explicitly with ICEs \citep{dahabreh2023efficient}.
For our context, relevant ICEs include (1) events occurring before the surrogate endpoint is measured that make the surrogate endpoint undefined, which would include death, or, for many applications, the target outcome; and 
(2) events occurring after the surrogate endpoint is measured that makes the target outcome undefined, which may include death.  For some disease contexts, allowing the target outcome to include death in a composite endpoint may address a relevant question, whereas for other contexts excluding deaths may yield a more desirable interpretation. For example, excluding death would be warranted if the disease under study has very low probability of causing death (much lower than the rate $Y=1$) such that most deaths are unrelated to the disease.

Studies of preventive vaccines illustrate a setting where target outcome occurrence before the surrogate endpoint is measured may be considered to render the surrogate endpoint ill-defined. For example, suppose the intent for the biomarker $S$ is to measure an antibody response induced solely by vaccination. 
Occurrence of the infectious disease target outcome would generate an antibody response that renders the measured $S$ reflecting a mixture of antibodies made by vaccination and by the infection, and often will be spiked high.  Including such participants would make it more difficult to model $g(X,S)$ and would make the target population a mixture of two quite distinct immunological groups. 

In settings where it is decided to exclude participants with $Y^{0-\tau}=1$, a TE target parameter that contrasts $E[Y(1)|Z=0,Y^{0-\tau}=0]$ and $E[Y(0)|Z=0,Y^{0-\tau}=0]$ is not a causal parameter due to different conditioning sets. 
One approach to recovering a causal parameter re-defines TE as a contrast of $E[Y(1)|Z=0,Y^{0-\tau}(1)=Y^{0-\tau}(0)=0]$ and 
$E[Y(0)|Z=0,Y^{0-\tau}(1)=Y^{0-\tau}(0)=0]$, which measures efficacy in the ``always-survivors" principal stratum who would be free of the target outcome by time $\tau$ under both treatment assignments (which implies imagining the assignment in the observational study). This approach has been commonly used for preventive vaccines (e.g., \cite{GilbertBletteShepherdHudgens2020}). 
In some other settings, including participants with $Y^{0-\tau}=1$ and ignoring target outcomes before $\tau$ might be acceptable, for example if the target outcome minimally affects $S$.

Appendix B discusses additional considerations for potentially relaxing our set-up of focus that only includes participants with $Y^{0-\tau}=0$ and hence does not accommodate the fact that in most real studies some participants will have
$Y^{0-\tau}=1$.

\subsection{Step 6: Quantify the uncertainty in the estimate of the statistical estimand}
\label{RoadmapStep6}

To meet the provisional approval bar of ``reasonably likely to infer sufficient treatment efficacy," the quantification of uncertainty in the estimation of TE should account for the spectrum of relevant uncertainty sources, including
(1) sampling variability in the estimation of $g(X,S)$ in the observational study; (2) sampling variability in $X$ and $S$ measured in the phase 3 study; 
(3) Margin for error due to the observational and phase 3 studies having different untreated population conditional distributions of $Y$ given $X$ and $S$; 
(4) Margin for error for an imperfect surrogate; and
(5) Margin for error due to any simplifications that are made in the handling of ICEs.

For the plug-in estimator of TE, uncertainty sources (1) and (2) can be accounted for by sandwich variance estimation or the bootstrap if a parametric or semiparametric model is used to estimate
$g(X,S)$, including if superlearner is used with not too overly-adaptive learners. That is, under all the identifiability conditions plus correctly specified models, the sandwich variance and the bootstrap provide asymptotically correct variance estimators for
$\widehat E[Y(0)|Z=0]$ and $\widehat E[Y(1)|Z=0]$, and hence for $\widehat{{\rm TE}}$.  The sandwich variance estimator for each
$\widehat E[Y(a)|Z=0]$ is derived using a stacked estimating equation \citep{stefanski2002calculus} that includes equations for each parameter in 
(\ref{eq: plug-in}), most of which are nuisance parameters. 
This estimating equations approach is detailed in Appendix C.
The bootstrap re-samples from both the observational study and the phase 3 study.

If superlearner with highly data-adaptive learners is used, then both standard sandwich variance estimation and 
the bootstrap do not provide asymptotically correct inferences.
Ways to remedy this include the highly-adaptive lasso (HAL) \citep{benkeser2016highly} or use of cross-fitting in the variance estimation (e.g., \cite{10.1214/aos/1176345863,robins2008higher,westling2023inference}).
Another option for nonparametric estimation/inference is
\citet{wolock2022framework}.

\comment{
\id{We could derive a fully efficient non-parametric estimator using the efficient influence function, and then use the EIF to compute variance estimates which are correct under assumptions. Happy to work on that if it is of interest.}

{\color{purple} PG 1-26-24: Great idea, and yes I'd appreciate that effort, it would contribute a lot.  Of course it would be good we first feel confident in the identifiability proof.}
}

For the one-step nonparametric efficient estimator
$\hat\theta_{a,\mbox{\text{\tiny one-step}}}$, for each $a$, under regularity conditions the variance can be consistently estimated by the empirical variance of $\phi_a(O_i,\hat\eta)$ across the $i=1,\cdots, n$ observations.
%

\subsubsection{Specification of the Bias Functions}
\label{biasfunctions}

Uncertainty sources (3) and (4) can be addressed by specifying the bias functions $u^{UC}(X,S)$ and $u^{CT}(X,S)$, respectively, that conservatively make estimates of TE smaller. A positive value of $u^{UC}(X,S)$ makes both the estimates of $E[Y(0)|Z=0]$ and $E[Y(1)|Z=0]$ smaller, such that both negative and positive values of $u^{UC}(X,S)$ could make TE smaller, such that conservative sensitivity analysis would need to consider a set of values both negative and positive.  In contrast, $u^{CT}(X,S)$ only affects $E[Y(1)|Z=0]$,
where a positive 
value of $u^{CT}(X,S)$ makes the estimate of $E[Y(1)|Z=0]$ larger and hence TE smaller.  Therefore a conservative analysis focuses on positive values of 
$u^{CT}(X,S)$. In conclusion, a conservative sensitivity analysis can be set up by defining maximum negative and maximum positive plausible values of 
$u^{UC}(X,S)$, and a maximum positive plausible value for 
$u^{CT}(X,S)$.  Including negative values of $u^{CT}(X,S)$ still may be of interest for a more complete sensitivity analysis, but positive values are salient for the provisional approval objective.

Once plausible sets are specified for the two bias functions, for every specific choice of $u^{UC}(X,S)$ and $u^{CT}(X,S)$ fixed, point estimates are obtained for 
$E[Y(1)|Z=0]$, $E[Y(0)|Z=0]$, and
TE, and the variance estimation accounting for uncertainty sources (1) and (2) yields a 95\% confidence interval for each of these three causal parameters.  Jointly accounting for uncertainty sources (1)--(4) can be achieved by reporting the range of point estimates (i.e., ignorance interval) and the union/envelope of 95\% confidence intervals for each causal parameter (i.e., 95\% estimated uncertainty interval \citep{Vansteelandtetal2006}). 
A success criteria guideline for provisional approval could be based on minimal bars for the left endpoint of the ignorance interval for TE and the left endpoint of the 95\% estimated uncertainty interval for TE.  

How can a plausible range for $u^{UC}(X,S)$ be specified?  It will commonly be the case that the observational studies are completed (or almost so) by the the time the phase 3 study collects data.  This means the untreated $A=0$ will not overlap in the two studies, implying it is not possible to control for secular calendar trends in target outcome incidence in Untreated-to-Control-transport.  Therefore, an idea for sensitivity analysis is to access external data bases to estimate population-level fluctuations in incidence over calendar time, and to use those results to guide specification of the sensitivity parameter $u^{UC}(X,S)$.  For example, if data are available on two observational studies conducted in different calendar periods of follow-up, then a linear regression could be fit for outcome $Y$ on $X,S$ and observational study, and $u^{UC}$ taken to be the point estimate (or confidence limit) of the coefficient for observational study. 
\citet{dahabreh2023sensitivity} focused on a constant bias function $u^{UC}(X,S) = u^{UC}$, and under this approach $u^{UC}$ could be specified to vary over a range that includes the plausible range of calendar fluctuations from the observational study to the phase 3 study.
Conduct of multiple observational studies in different geographic regions/populations would aid specification of $u^{UC}$.

Another idea 
considers the similarity of the relationship between $X$ and $S$
in the observational study vs. the phase 3 placebo arm, which can be studied with reasonable precision.
Under the premise that $P(X,S|Z=0,A=0) \neq P(X,S|Z=1,Z=0)$ would likely imply
$P(Y|X,Z=0,A=0,S) \neq P(Y|X,Z=1,A=0,S)$, the greater the dissimilarity 
in the $(X,S)$ relationship
the more need to increase the bias function $u^{UC}(X,S)$.

How can a plausible range for $u^{CT}(X,S)$ be specified?  This is challenging given the absence of any previous phase 3 randomized trials to help validate the surrogate, and we have emphasized that in our problem context there are too-few $Y=1$ outcomes in any randomized trial to estimate $u^{CT}(X,S)$. However, we know $S$ provides a (very) strong correlate of risk, or otherwise the pathway for provisional approval would have little chance of success, and this strong association provides some likelihood that the marker has at least some partial validity as a surrogate endpoint.  One simple recipe is as follows. Consider $X$-specific treatment efficacy on the 
additive difference scale: ${\rm TE}^{ATE}(X) = E[Y(1)|X,Z=0] - E[Y(0)|X,Z=0]$. Then the proportion of the treatment effect on the target outcome explained by $S$ (original
version of \citet{Freedman1992}) can be written as
$${\rm PTE}(X) := 1 - \frac{u^{CT}(X,S)}{{\rm TE}^{ATE}(X)}.$$

\noindent Therefore 
\begin{eqnarray}
    u^{CT}(X,S) = {\rm TE}^{ATE}(X) \left[ 1- {\rm PTE}(X) \right]. \label{eq: PTEspec}
\end{eqnarray}

\noindent Set multiplicative treatment efficacy ${\rm TE}(X) = 1 - E[Y(1)|X,Z=0] \slash E[Y(0)|X,Z=0]$ to the preferred target product profile (TPP) level for the candidate treatment that indicates clear merit for use.
Next, set a lower bound for PTE($X$),
such as 0.5 \citep{lin1997estimating}, which may be credible given the high correlation of $S$ with $Y$ in the observational study as noted above. 
Next, as 
${\rm TE}^{ATE}(X) = -P(Y(0)=1|X,Z=0)$TE($X$), the last parameter to specify is the placebo arm disease rate, with one choice being the point estimate of
$P(Y(0)=1|X,Z=0)$ in the observational study.

For example, suppose $P(Y(0)=1|X,Z=0)$ is set to 0.005 based on the exercise to set a plausible range for $U^{UC}(X,S) = U^{UC}$, and TE is set to 0.7, such that
${\rm TE}^{ATE} = -0.0035$.  Then, setting PTE($X$) = 0.5 yields
$u^{CT}(X,S) = -0.00175$, and accordingly $u^{CT}(X,S) = u^{CT}$ could be varied over the range -0.00175 to 0.00175, and the ignorance interval and 95\% EUI could be calculated based on these two extreme values.  

Another consideration for specifying $u^{CT}(X,S)$ is that the more unmeasured confounding of the effect of $S$ on $Y$ in the estimation 
of $g(X,S)$, the more opportunity for Control-to-Treated-transport to fail.  Therefore, domain knowledge on the confounders included in $X$, as well as concerns about missing confounders, influence the range of specified $u^{CT}(X,S)$.

To address additional uncertainty sources such as (5), additional sensitivity parameters could be specified and varied over plausible ranges and their variability accounted for in the ignorance intervals and estimated uncertainty intervals.  Where these elaborations are formidable, a simpler approach looks for an ignorance interval and 95\% estimated uncertainty interval [accounting for (3), (4) but not (5)] that sail well over the pre-specified success bars, providing some assurance in the presence of difficult-to-fully-quantify uncertainty.


\subsection{Step 7: Compare feasible complete analytic designs (Steps 1--6) using outcome-blind simulations}

We phrase this step in the same way as \citet{dang2023causal} that described the causal roadmap for generating high-quality real-world evidence.  The statistical analysis plan for the phase 3 study needs to be completed before availability of the phase 3 data.  However, it may be possible to conduct outcome-blind simulations of the observational study and hypothetical phase 3 study to fine-tune the choices that must be made, including on: (1) how to estimate $g(X,S)$ including which variables
$X$ and $S$ to include; (2) how to specify the bias functions $u^{UC}(X,S)$ and
$u^{CT}(X,S)$; and (3) how to ensure that EUIs for the causal parameter of interest adequately capture uncertainty in both the observational study and phase 3 study.  These questions are considered in the simulation study in Section \ref{simtudyGBS}.
Iterative communication of the simulation results with stakeholders aids finalization of choices.

\subsection{Comparison of the present work to Athey et al. (2024)}
\label{diffarticles}

\begin{enumerate}
\item The two articles use similar notation except our study indicator $Z$ is their $\cal P$ and our binary treatment $A$ is their $W$.  In the following notes we use the notation of the present article.


\item The two articles consider the same target causal parameter of interest -- TE -- a contrast in $E[Y(1)|Z=0]$ vs. $E[Y(0)|Z=0]$. This work develops all elements allowing for a general contrast function $h(x,y)$ satisfying $h(x,y)=0$ if and only if $x=y$, given that a multiplicative contrast such as $h(x,y) = $ log$(E[Y(1)|Z=0],E[Y(0)|Z=0])$ or $h(x,y) = 1 - E[Y(1)|Z=0] \slash E[Y(0)|Z=0]$ is needed for our provisional approval application, whereas Athey et al. restrict attention to an additive difference contrast $h(x,y)=E[Y(1)|Z=0] - E[Y(0)|Z=0]$. This is a non-substantive difference inasmuch as the Athey et al. results could be readily generalized to handle a general contrast.

\item This work supposes all observational study participants have treatment level $A=0$ known whereas Athey et al. assume $A$ is missing/unknown. An implication is Athey et al.'s Assumption 1 of a single random sample is slightly different from our set-up, in this one specific element. Under both the Comparability Assumption and the (perfect) Surrogacy Assumption noted below, this difference does not affect the identifiability results nor the estimators, such that results of the two articles are equivalent in this case.  Under violations of either of these assumptions, the identifiability results and hence also the estimators are different for the two articles.

\item Related to the previous point, both articles provide a nonparametric efficient influence function for use in estimation, which are equal under both the Comparability Assumption and Surrogacy Assumption and differ otherwise due to the different set-up.  

\item Both articles use the ``optimal surrogate" (our nomenclature from \citet{PriceGilbertvanderLaan2018}) or equivalently ``surrogate index" (Athey et al. nomenclature) as a central ingredient of the results: $E[Y|X,Z=0,A=a,S]$. Indeed, a significant idea common to both articles is to make use of a conditional regression that can depend on multivariable $S$, an idea that we previously championed in \citet{PriceGilbertvanderLaan2018} and picked up in this work. 
Athey et al. provide a valuable summary of the benefits of multiple surrogates in their Section 3.2.3.

\item Both articles assume that in the experimental study $Z=0$, treatment assignment $A$ is unconfounded (strong ignorability), equivalently the $Z=0$ study is randomized within levels of baseline characteristics $X$.  This assumption is A2 in this work and Assumption 2 in Athey et al.   

\item Both articles make a Comparability Assumption, in different ways. Our assumption A4 equates two conditional means offset by the Untreated-to-Control transport bias function $u^{UC}(X,S)$, and Athey et al.'s Assumption 4 expresses conditional independence.  

\item Both articles use a Surrogacy Assumption (a Prentice valid surrogate: our assumption A6; Athey et al. Assumption 3) as a key assumption for the results, with difference that our assumption is expressed in terms of equal conditional means and the Athey et al. assumption is expressed as conditional independence.  
For identifiability of the target causal parameter of interest, the assumption of equal conditional means suffices, with conditional independence not necessary; we speculate that it was not Athey et al.'s goal to define minimally sufficient identifiability assumptions as they had other reasons to prefer to use the stronger conditional independence assumption.
In addition, the present article does not state the Surrogacy Assumption apart from a bias function given that for the provisional approval application the data analysis prioritizes scenarios with a non-zero bias function.

\item Elaborating on the last point, the two articles make a different ``style choice" regarding
how to include bias functions in relation to the key Surrogate Assumption and Comparability Assumption. Athey et al. introduces these key assumptions under the ideal case that both assumptions hold, and provide results expressing the bias that results from deviations from these assumptions. The present work 
includes the bias functions as fixed and user-specified functions directly in the identifiability assumptions. Our provisional approval application drove this choice, where for this application data analysis assuming bias is most germane based on our understanding that regulators will generally require explicit conservative inference to provide sufficient evidence undergirding a provisional approval decision, where the pre-specified success benchmark in the statistical analysis plan would include non-zero bias functions.

\item Athey et al. included three representations of the statistical estimand of interest [their equation (4.1) and our equation (\ref{eq: Untreatedtransport})] using a surrogate index, a surrogate score [their Definition 2: surrogate score $P(A=1|S=s,X=x,Z=0)$], or both, with utility that each representation indicates different parameters that require estimation and for some applications one or another representation may be more advantageous. These representations are equations (4.2), (4.3), (4.4) in Athey et al. Our article focuses on a representation most similar to (4.2), and in the special case that the Comparability Assumption and the Surrogacy Assumption hold, it is straightforward to write our statistical estimand in any of the three representations. 



\item This work emphasizes the fact that the surrogates $S$ are measured via a two-phase sampling design in both studies, developing all results accounting for this ubiquitous data reality for the provisional approval application. Athey et al. considers complete data on the surrogates $S$ in the observational study; this is a minor difference from the perspective that all results of Athey et al. could be extended to account for two-phase sampling of $S$ under a missing at random assumption. Yet it would be a non-trivial amount of work to extend the identifiability and estimation results, and there are many ways that the estimators could be extended to account for the missing data on $S$.

\item This work considered the issues of the target outcome $Y = I(T \le t_0)$ being subject to right-censoring and considered Intercurrent events (ICEs).

\end{enumerate}

\section{Case Study: Application to Group B Streptococcus Vaccination}

For many years the GBS vaccine field has been pursuing a provisional approval pathway based on antibody markers measured in infant cord blood [IgG antibody levels to capsular or Alpha proteins inside of vaccine candidates, and 
Opsonophagocytosis Killing Assay (OpKA) readouts] that have been shown to strongly inversely correlate with IGbsD in natural history studies
[e.g., \citet{madhi2021association,madhi2023potential,dangor2023association}].
Vaccine developers are currently pursuing a development pathway that learns about a surrogate endpoint from multiple observational studies and then estimates vaccine efficacy in a phase 3 target trial based on this surrogate. 
The surrogate endpoint is derived from 
antibody markers measured in cord blood: IgG levels against each GBS antigen represented in the vaccine construct antigens and a weighted average of these levels (``average IgG"), and OpKA levels against each of the same GBS antigens plus a weighted average of these levels (``average OpKA").

The sero-epidemiological observational studies, all currently ongoing, are the COP01-WITS/GBS Alpha study in South Africa (led by professor Shabir Madhi), 
the COP02-SGUL/iGBS3 study in the United Kingdom (led by professor Kirsty Le Doare), 
and the EDCTP-sponsored PREPARE study
in Denmark, France, Italy, Malawi, the Netherlands, South Africa, Uganda, United Kingdom, United States (led by professor Kirsty Le Doare), which are similar to the 
GBS-CoP and V98\_280BTP observational 
studies in South Africa that were recently published \citep{madhi2021association,madhi2023potential}. Each study collects cord blood samples and/or acute-illness samples from infant IGbsD cases from 0 to 90 days of age and cord blood samples from non-cases/controls, enabling analyses to assess the association of cord-blood antibody markers with IGbsD.  

Phase 3 GBS vaccine trials in planning randomize pregnant mothers to receive two vaccine or placebo doses 
starting in the third trimester,
with cord-blood collected from all infants for potential measurement of the set of antibody markers $S$ (hence time $\tau$ is the birth/delivery visit). The antibody markers will be measured from a random sample of live-born infants of vaccinated and placebo mothers, oversampling infants born to GBS colonized mothers and including infants born to GBS uncolonized mothers.
The visit schedule and sampling of cord blood is harmonized across the multiple observational studies and the phase 3 study, and the same antibody markers are measured using the same instruments and protocol. 
The primary analysis estimates vaccine efficacy against IGbsD 
for the phase 3 sub-population of live-born infants with GBS colonized mothers based on the surrogate endpoint that is learned (i.e., an estimate of $g(X,S)$) 
from data analysis of the observational studies pooled.  Section \ref{ICEsec} discusses how this analysis that makes inferences for colonized mother-infant dyads relates to the objective to draw conservative inferences about vaccine efficacy for the whole randomized phase 3 population.

Below we consider specific application of the Causal Roadmap steps outlined in the previous section. We restrict to Steps 1--2, 5--7 for which the GBS vaccine context impacts considerations. 

\subsection{Step 1: Specify the causal model based on available knowledge of the context and proposed study}

The concentration of IgG antibodies that bind to various GBS surface proteins is highly predictive of IGbsD in many natural history studies.  The OpKA functional assay is believed to measure a causal mechanism of vaccine protection.  Thus, some subject matter experts suggest the perfect-surrogate causal model of Figure \ref{fig:dag_phase3} [Panel (A)] approximately holds.  However, one reason the imperfect-surrogate causal model [Panel (B)] is more appropriate is that IgG and OpKA levels for vaccinated and not infected with GBS vs. naturally GBS infected mothers may have different time-patterns in the infant from pre-birth through 90 days of age. A second reason is that the vaccine exposes a mother and baby to a specific GBS protein whereas natural-infection exposes a mother and baby to the entire GBS bacterium, meaning that natural-infection induces additional immune responses, and, if these responses help protect against IGbsD, then the bias function $u^{CT}(X,S)$ would likely be positive.  

\subsection{Step 2: Define the causal parameter of interest}
The causal parameter of interest is VE = $1 - E[Y(1)|Z=0]\slash E[Y(0)|Z=0]$ with $Y=I(T \leq 90)$: vaccine efficacy against IGbsD occurrence through $t_0=90$ days of life in the population of live-born infants with GBS colonized mothers.

%
%
%

\subsection{Step 5: Estimate the statistical estimand}
\label{GBSstep5}

We consider the plug-in estimator and the one-step estimator of 
$E[Y(0)|Z=0]$, $E[Y(1)|Z=0]$, and the target parameter TE = VE.
 For estimating $g(X,S)$ from the sero-epidemiological observational studies, the baseline variables $X$ are taken to be a set of known prognostic risk factors for IGbsD: 
$X_1$ the indicator of gestational age less than 37 weeks, and 
$X_2$ maternal age in years where younger age is a risk factor.  The putative surrogate $S$ is the average log$_{10}$ 
IgG concentration against the four GBS alpha types in the vaccine.
As noted in Section \ref{engine}, many different IPS-weighted regression estimators for $g(X,S)$ may be considered.
A superlearner estimator has appeal for providing an estimated optimal 
surrogate (EOS) to use in the estimation formulas, where several EOSs would be developed each under a different set of input variables $(X,S)$, providing an empirical approach to selecting a best-predictive and parsimonious $(X,S)$ to include in the phase 3 surrogate endpoint study data analysis \citep{PriceGilbertvanderLaan2018}. For specification of the phase 3 statistical analysis plan, it would be useful to conduct a simulation study that compares performance of different implementations of the estimators (and variance estimators), 
not only considering different input variable sets but also different libraries of learners in the superlearner and different implementation details such as loss function and cross-validation scheme.

\subsubsection{Intercurrent events}
\label{ICEcolonized}

The set of eligible participants for inclusion in the sero-epidemiological study data analysis are live-born infants of mothers colonized with GBS. As such, the methods provide estimation of VE for this colonized population.
However, the public health goal for the indication of the vaccine is to include all pregnant persons irrespective of GBS colonization, avoiding the need for GBS colonization screening prior to vaccination.
Therefore, an important goal is to infer that VE for the entire randomized cohort of pregnant mothers is sufficiently high. 
If the probability of IGbsD is zero for infants born to uncolonized mothers, then this inference is fairly straightforward.  Let TE$^{\rm Colon.}$ be treatment efficacy against the target outcome in the colonized subgroup and let TE.against.Colon. be treatment efficacy against maternal GBS colonization by time $\tau$, i.e., TE.against.Colon. $= 1 - P(C(1)=1|Z=0) \slash P(C(0)=1|Z=0)$ with $C(a)$ the potential outcome indicator of whether the mother is colonized by $\tau$ under randomization assignment $a$, for $a=0,1$.  Then 
\begin{eqnarray}
{\rm TE}= 1  - (1 - {\rm TE}^{\rm Colon.})(1 - {\rm TE.against.Colon.}), \label{eq: VEcolonizedform}
\end{eqnarray}
\noindent which implies that as long as the vaccine does not increase the rate of colonization, then vaccine efficacy in the randomized cohort will be at least as high as that in the colonized. Vaccine efficacy against GBS colonization can be readily estimated in the phase 3 study, such that the above formula can be used to seek assurance that focusing on estimating TE in the colonized supports sufficient treatment efficacy for the whole randomized cohort.

Indeed, the risk of IGbsD in uncolonized mothers is close to zero \citep{vekemans2019maternal} (some would argue equal to zero), which is the reason the sero-epidemiological studies restrict the study cohort to colonized mothers. In fact, being uncolonized means that the bacteria are not alive and growing in which case IGbsD should be impossible; however, colonization is measured by a diagnostic procedure applied at one or a discrete set of sampling time points, and imperfect sensitivity of the colonization diagnostic test could imply low residual risk of IGbsD in infants born from mothers diagnosed as uncolonized. 
Missing detection of colonized GBS could also occur due to post partum GBS acquisition and non-vertical GBS transmission to the infant.

If there is concern of non-negligible risk of IGbsD for infants born to mothers diagnosed as uncolonized, then an approach to inferring sufficient VE for the whole randomized cohort is to rely on the estimate of VE for the colonized sub-population, supplemented by arguments that this estimate of VE serves as a lower bound of the estimate of VE for the whole randomized cohort. One source of evidence would be data showing that the GBS vaccine reduces the rate of colonization in the randomized cohort, or at least does not increase the rate, as noted above. This can readily be studied as an objective in the phase 3 trial. A second source of evidence would be data supporting that VE against IGbsD in uncolonized mothers is at least as high as VE against IGbsD in colonized mothers. In principle, argumentation for this point could be based on the sampling design for measurement of $S$ in the phase 3 study, measuring $S$ for both colonized and uncolonized participants, from both the vaccine and placebo arms, and applying the VE estimation formula to both subgroups.  However, because the estimator $\hat g(X,S)$ would be different for the colonized vs. uncolonized groups, this approach would only be viable if the observational study includes the uncolonized group in the correlates of risk model building. This is not possible for the real GBS studies given zero probability of sampling $S$ for the uncolonized group.  However, some less direct evidence could be generated from the phase 3 trial in the form of comparing the distribution of $S$ between the colonized vs. uncolonized groups within each of the vaccine and placebo arms.  
The assumption of vaccine efficacy being as least as high in uncolonized mothers may be reasonable based on the observation that vaccines typically provide better efficacy when the amount of pathogen exposure (e.g., number of virus copies is less) \citep{kaslow2021force}. 
Counterbalancing this argument, an immunological reason why the colonized could have have higher efficacy is that colonization generates immune responses and the vaccination serves as a boost, whereas the uncolonized may be naive to the pathogen such that the vaccination is like a prime, and typically vaccines confer greater efficacy in groups with prior infection compared to in naive groups (e.g., \cite{Sadoffetal2022}).

\comment{
If the vaccine lowers the rate of stillbirth or the rate of GBS colonization, as it is hypothesized to do \citep{vekemans2019role}, then the fact the analysis conditions on live-born infants of colonized mothers could cause post-randomization selection bias
when defining TE as a contrast in $E[Y(0)|Z=0,Y^{0-\tau}=0]$ vs. $E[Y(0)|Z=0,Y^{0-\tau}=0]$ in this eligible group. 
}
In the GBS application, $Y^{0-\tau}$ is the indicator of acquiring IGbsD by birth, such that the simplifying assumption $P(Y^{0-\tau}=1)=0$ amounts to no in-utero IGbsD that has onset on the date of birth.
If such events do occur, then additional investigations would be warranted to provide evidence for whether restricting the analysis to infants with $Y^{0-\tau}=0$ would be expected to affect VE in the whole cohort.  An alternative framing would redefine 
$Y^{0-\tau}=1$ to mean IGbsD onset {\it before} birth and therefore structurally to make the assumption 
$P(Y^{0-\tau}=1)=0$ hold, in which case $S$ measured from cord blood would need to be considered definable, where some infants may have IGbsD onset at birth that are assigned a failure time of one day.  This latter framing may be reasonable based on the fact that infants receive the $S$ immune marker levels passively from the mother.
%

\subsection{Step 6: Quantify the uncertainty in the estimate of the statistical estimand}

It is of interest to compare the two variance estimation approaches for the plug-in estimator (sandwich, bootstrap)
 for the specific context of the planned/available GBS sero-epidemiological studies and potential phase 3 study designs. 
We evaluate these two variance estimators in Section \ref{RoadmapStep6}.  It is also of interest to study the strategies for specifying the bias functions considered in Section \ref{biasfunctions}.  

\comment{
Note that the bootstrap is only known to be valid if the set of learners used in the super learner for covariate-adjustment in TMLE are ``smooth and not overly data-adaptive" satisfying Donsker conditions (e.g., parametric regression models like generalized additive models, lasso, HAL TMLE \citep{benkeser2016highly}), but for more data-adaptive algorithms the confidence intervals may not have correct asymptotic coverage.  This is repaired by cross-fitting, which would require additional programming.  Because of this issue, the current implementation will only include learners that satisfy Donsker conditions.  If more data-adaptive learners are deemed needed, later on the code can be updated to add cross-fitting.
}



\subsection{Step 7: Simulation Study to Plan for VE Estimation in the GBS Vaccine Phase 3 Study}  
\label{simtudyGBS}


 We generate data for mother-infant (live born) dyads observed in a sero-epidemiological study or enrolled in the phase 3 placebo-controlled vaccine trial. We designed simulation conditions in the sero-epidemiological study to roughly match published GBS characteristics for colonized mothers. In particular, we specify probability of IGbsD by 90 days at 0.005 \citep{vekemans2019maternal}
 and geometric mean cord-blood IgG concentration of 0.01 and 0.04 in IGbsD cases and controls (observed free of IGbsD through 90 days of age), respectively \citep{dangor2023association}. Additionally,  we consider $X=(X_1,X_2,X_3)$ with $X_1$ and $X_2$ defined in Section \ref{GBSstep5}, and we add a continuous noise variable $X_3$ unrelated to outcome (an accidentally adjusted for variable). For the phase 3 study, the same variables were simulated. Appendix D provides justification for the simulation conditions.
 
Two simulation studies are conducted. First, data are simulated with both bias functions truly equal to zero, and 
we evaluate bias, standard error, and coverage of 95\% confidence intervals for each $E[Y(a)|Z=0]$ and VE when conducting the analysis with the bias functions set to zero. The purpose of this simulation study is to verify correct properties of estimator performance under known conditions.
Secondly, we conduct the analysis using specified bias functions to make inferences on VE conservative. The purpose of these simulations is to evaluate power to meet the success criteria (defined as 95\% EUI for VE $\ge 0.3$) for realistic methods' implementations that requires non-zero bias functions.  The simulations focus on evaluation of the plug-in estimator (evaluation of the one-step estimator underway).

\subsubsection{Simulation Study 1: Ideal conditions with zero bias functions and no sensitivity analysis}

The purpose of the first simulation study is to verify correct operating characteristics (bias, coverage probability of confidence intervals, correct variance estimation) of the plug-in estimation approach.
 We generate data under the conditions that the true bias functions $u^{UC}(X,S)$ and $u^{CT}(X,S)$ are zero. We simulate data for 39,000 participants in the sero-epidemiological study ($Z=1, A=0$). We simulate data for 6,200 participants enrolled in the phase 3 vaccine trial ($Z = 0$), with 3,100 each randomized to the vaccine ($A=1$) and placebo ($A=0$) arm.

We fix the following simulation conditions to be the same in the two studies $Z= \{0, 1\}$: 

\begin{enumerate}
    \item $X = (X_1, X_2, X_3)$ where $X_1 | Z \sim \text{Bernoulli}(0.05)$, $X_2|Z \sim \text{Uniform}(18,40)$, and $X_3 | Z \sim \text{Normal}(0,1)$
    \item $S | X, A = 0, Z \sim \text{Normal}(-1.45, 0.0225)$
    \item $Y | X, A, Z, S \sim$ Bernoulli with $P(Y = 1 |A = 0, Z, S, X_1, X_2, X_3) = -17.1 - 8.2 S + 0.69 X_1 - 0.03 X_2$
    \item $Y = I(T \le t_0)$ is always observed (for simplicity); i.e. no right-censoring by 90 days of age
\end{enumerate}


In order to induce $u^{CT}(X,S) = 0$, we set the conditional distribution function of $Y$ equal for the vaccine and placebo arms: $P(Y = 1 |A = 1, Z = 0, S, X_1, X_2, X_3) = P(Y = 1 |A = 0, Z = 0, S, X_1, X_2, X_3)$.

To generate VE $= \{ 0, 0.5, 0.9\}$, we manipulate the distribution of the biomarker in the vaccine arm $S|A = 1, Z = 0$:
\begin{itemize}
    \item To create VE $= 0$, set $S|A =1, Z = 0 \sim \text{Normal}(-1.45, 0.0225)$
    \item To create VE $= 0.5$, set $S|A =1, Z = 0 \sim \text{Normal}(-1.296, 0.04)$
    \item To create VE $= 0.9$, set $S|A =1, Z = 0 \sim \text{Normal}(-1.08, 0.0441)$
\end{itemize}


Finally, we set a case-control sampling design for measuring $S$ in cord blood in the observational study as follows: we sample $S$ from all IGbsD cases and a simple random sample of controls with the number set to 5 times the number of cases. In the phase 3 trial, we sample $S$ from a simple random sample of 100, 250, or 500 infant participants in each randomization arm. 

For estimation and inference, we first estimate $g(X,S)$ with IPS-weighted logistic regression of $Y$ on $(X, S)$ including participants with $S$ measured in the observational study. The IPS weight $\pi(X,Z,A,\tilde T,\Delta)$ is set to the true sampling weights: 1 for cases and $n_{controls} \slash 5n_{cases}$ for the controls, where $n_{cases}$ is the number of participants with $Y=1$ and $n_{controls}$ is the number of participants with $Y=0$. To obtain an estimate for the expectation in the summand of (\ref{eq: plug-in}), we then regress the fitted values $\hat{g}(X, Z=1, S)$ on $X$, in the phase 3 study ($Z = 0$) in each arm using IPS-weighted linear regression.

We then use the plug-in estimator equation (\ref{eq: plug-in}) to calculate $\hat{E}[Y(0)|Z=0]$ and $\hat{E}[Y(1)|Z=0]$, yielding $\log(1 - \widehat{{\rm VE}}) = \log \lbrace 
\hat{E}[Y(1)|Z=0] \slash \hat{E}[Y(0)|Z=0] \rbrace$. We calculate bootstrap and sandwich standard errors for $\hat{E}[Y(0)|Z=0]$, $\hat{E}[Y(1)|Z=0]$, and $\log(1 - \widehat{{\rm VE}})$. The bootstrap re-samples cases and controls separately in the observational data set, and vaccine and placebo recipients separately in the phase 3 data set. The sandwich error estimation strategy is described in Appendix C. From these standard errors, we obtain Wald confidence intervals. To estimate $\widehat{{\rm VE}}$ and its corresponding confidence intervals, we transform the point and confidence interval estimates of $\log(1 - \widehat{{\rm VE}})$. 

Across 800 simulations, we report bias and confidence interval coverage for $E[Y(0)|Z=0]$, $E[Y(1)|Z=0]$, and VE. Additionally, we report bootstrap standard error, sandwich standard error, and Monte Carlo empirical standard deviations for $\hat{E}[Y(0)|Z=0]$, $\hat{E}[Y(1)|Z=0]$, and $\log(1 - \widehat{{\rm VE}})$. 

\subsubsection{Simulation Study 1: Results}

\begin{figure}[h!]
    \centering
    \includegraphics[width=0.7\textwidth]{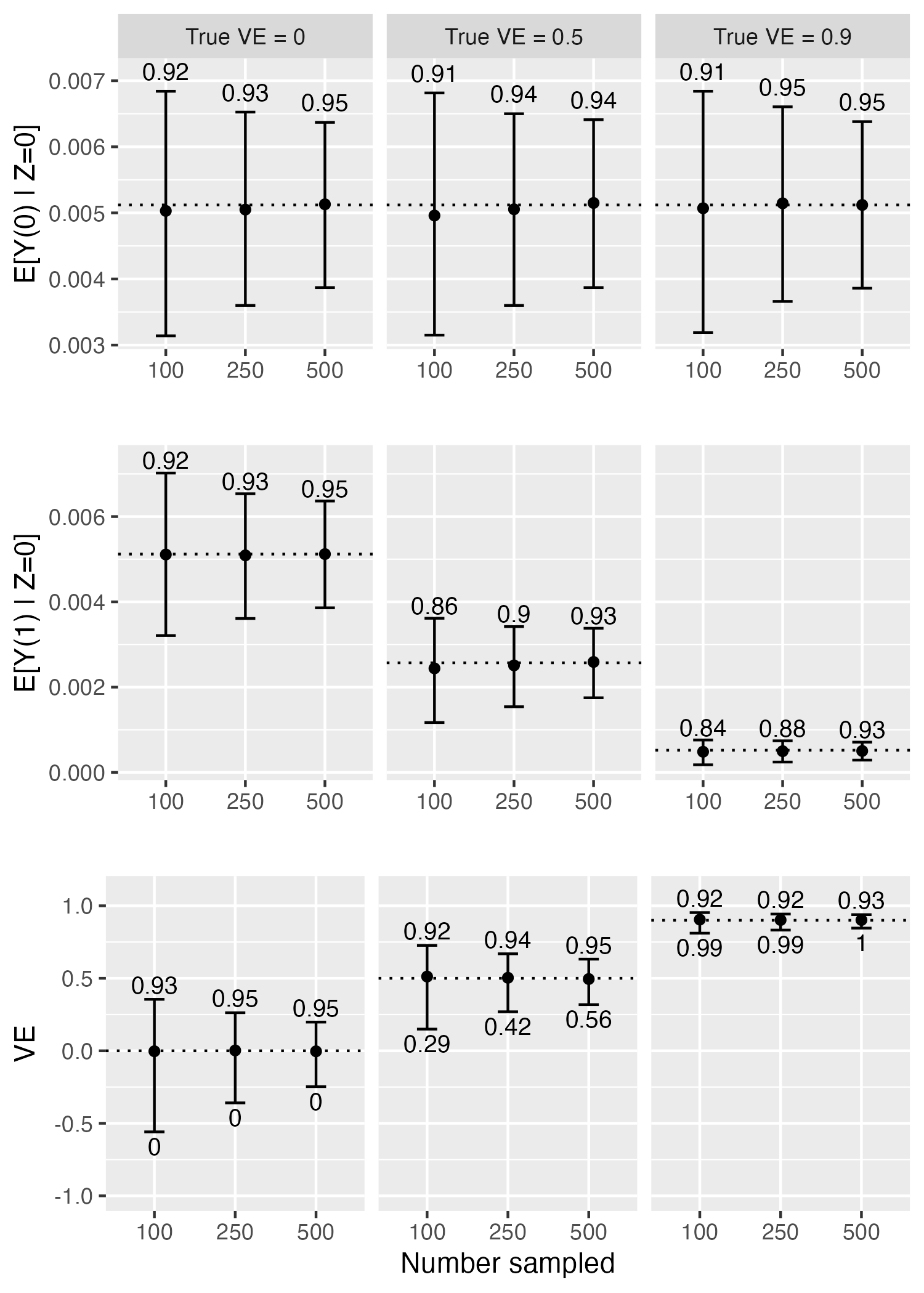}
    \caption{Results for Simulation Study 1 across true VE = 0, 0.5, and 0.9. The top, middle, and bottom panels display results for estimating $E[Y(0)|Z=0]$, $E[Y(1)|Z=0]$, and VE, respectively. Each graph depicts  results across 100, 250, and 500 with $S$ data in each of the vaccine and placebo arms of the phase 3 study $Z=0$. Points represent the median estimate across simulations, and error bars display the median  lower and upper sandwich 95\% confidence interval bounds. Dotted horizontal lines are placed at the true value. The numbers above each error bar display the confidence interval coverage, and the numbers below the error bar in the VE panel show power to meet the success criteria (defined as 95\% EUI for VE $\ge 0.3$).}
    \label{fig:sim_res1}
\end{figure}

Results for Simulation Study 1 are shown in Figure \ref{fig:sim_res1} and Table \ref{tab:sim_res1}. For all three simulation settings, we observed low bias for $\hat{E}[Y(0)|Z=0]$, $\hat{E}[Y(1)|Z=0]$, and $\widehat{{\rm VE}}$. Standard errors decreased and confidence interval coverage increased for $\hat{E}[Y(0)|Z=0]$, $\hat{E}[Y(1)|Z=0]$, and $1 - \log(\widehat{{\rm VE}})$ across increasing numbers sampled. Approximately nominal coverage of 95\% confidence intervals was obtained for $E[Y(0)|Z=0]$, $E[Y(1)|Z=0]$, and VE in the setting where 500 participants were sampled for each treatment arm of the $Z=0$ study. We generally observed mild discrepancies between the bootstrap and sandwich standard errors; the bootstrap standard errors for VE tended to be slightly larger. Empirical standard deviations tended to be higher than both estimated standard errors, with larger differences in settings where fewer participants were sampled and with the highest true VE. We believe this is due to the rare event setting with a limited number of $Y=1$ outcomes, as limited statistical information is available through the low number of cases. Additional simulations increasing the failure rate by about 3-fold 
showed closer agreement between the bootstrap, sandwich, and Monte Carlo standard errors (Appendix D, Section \ref{simresultssupp}).

\begin{table}[]
\caption{Results for Simulation Study 1 on empirical bias, average standard error (sandwich and bootstrap), standard deviation of estimates, and 95\% confidence interval coverage for $E[Y(0)|Z=0], E[Y(1)|Z=0]$. SE (bs) = bootstrap standard error, SE (sw) =  sandwich standard error, SD = standard deviation of estimates, Cov = 95\% confidence interval coverage using sandwich standard error, Sampled = number with $S$ data per arm in phase 3 study. The standard errors for $\log (1-\widehat{{\rm VE}})$ are displayed for VE.}
\begin{tabular}{llllll}
\rowcolor[HTML]{C0C0C0} 
              & {\color[HTML]{333333} Bias} & {\color[HTML]{333333} SE (bs)} & {\color[HTML]{333333} SE (sw)} & {\color[HTML]{333333} SD} & {\color[HTML]{333333} Cov} \\
\rowcolor[HTML]{B7B7B7} 
True VE = 0   &                             &                                &                                &                           &                            \\
\rowcolor[HTML]{EFEFEF} 
Sampled = 100 &                             &                                &                                &                           &                            \\
$E[Y(0)|Z=0]$          & -9.00E-05                   & 0.000948                       & 0.000914                       & 0.00129                   & 0.92                       \\
$E[Y(1)|Z=0]$          & -1.00E-05                   & 0.000978                       & 0.000935                       & 0.00113                   & 0.92                       \\
VE            & -0.00322                    & 0.25                           & 0.229                          & 0.274                     & 0.93                       \\
\rowcolor[HTML]{EFEFEF} 
Sampled = 250 &                             &                                &                                &                           &                            \\
$E[Y(0)|Z=0]$          & -7.00E-05                   & 0.000702                       & 0.000721                       & 0.000811                  & 0.93                       \\
$E[Y(1)|Z=0]$          & -3.00E-05                   & 0.000698                       & 0.00072                        & 0.000764                  & 0.93                       \\
VE            & 0.00254                     & 0.167                          & 0.156                          & 0.167                     & 0.95                       \\
\rowcolor[HTML]{EFEFEF} 
Sampled = 500 &                             &                                &                                &                           &                            \\
$E[Y(0)|Z=0]$          & 1.00E-05                    & 0.000587                       & 0.000637                       & 0.000672                  & 0.95                       \\
$E[Y(1)|Z=0]$          & 0                           & 0.000587                       & 0.000632                       & 0.000692                  & 0.95                       \\
VE            & -0.00354                    & 0.123                          & 0.117                          & 0.13                      & 0.95                       \\
\rowcolor[HTML]{B7B7B7} 
True VE = 0.5 &                             &                                &                                &                           &                            \\
\rowcolor[HTML]{EFEFEF} 
Sampled = 100 &                             &                                &                                &                           &                            \\
$E[Y(0)|Z=0]$         & -0.00016                    & 0.000935                       & 0.000896                       & 0.00107                   & 0.91                       \\
$E[Y(1)|Z=0]$          & -0.00013                    & 0.00061                        & 0.000573                       & 0.001                     & 0.86                       \\
VE            & 0.014                       & 0.312                          & 0.288                          & 0.365                     & 0.92                       \\
\rowcolor[HTML]{EFEFEF} 
Sampled = 250 &                             &                                &                                &                           &                            \\
$E[Y(0)|Z=0]$          & -6.50E-05                   & 0.000708                       & 0.000721                       & 0.000786                  & 0.94                       \\
$E[Y(1)|Z=0]$          & -6.00E-05                   & 0.000472                       & 0.00046                        & 0.000593                  & 0.9                        \\
VE            & 0.0055                      & 0.218                          & 0.205                          & 0.24                      & 0.94                       \\
\rowcolor[HTML]{EFEFEF} 
Sampled = 500 &                             &                                &                                &                           &                            \\
$E[Y(0)|Z=0]$         & 3.00E-05                    & 0.000588                       & 0.000634                       & 0.000667                  & 0.94                       \\
$E[Y(1)|Z=0]$         & 2.00E-05                    & 0.000399                       & 4.00E-04                       & 0.000452                  & 0.93                       \\
VE            & -0.003                      & 0.166                          & 0.156                          & 0.169                     & 0.95                       \\
\rowcolor[HTML]{B7B7B7} 
True VE = 0.9 &                             &                                &                                &                           &                            \\
\rowcolor[HTML]{EFEFEF} 
Sampled = 100 &                             &                                &                                &                           &                            \\
$E[Y(0)|Z=0]$          & -5.00E-05                   & 0.000938                       & 0.000902                       & 0.00121                   & 0.91                       \\
$E[Y(1)|Z=0]$          & -3.60E-05                   & 0.000142                       & 0.00014                        & 0.000235                  & 0.84                       \\
VE            & 0.007                       & 0.38                           & 0.35                           & 0.433                     & 0.92                       \\
\rowcolor[HTML]{EFEFEF} 
Sampled = 250 &                             &                                &                                &                           &                            \\
$E[Y(0)|Z=0]$          & 2.50E-05                    & 0.00071                        & 0.00073                        & 0.000793                  & 0.95                       \\
$E[Y(1)|Z=0]$          & -2.25E-05                   & 0.000118                       & 0.00012                        & 0.000177                  & 0.88                       \\
VE            & 0.004                       & 0.297                          & 0.278                          & 0.344                     & 0.92                       \\
\rowcolor[HTML]{EFEFEF} 
Sampled = 500 &                             &                                &                                &                           &                            \\
$E[Y(0)|Z=0]$           & 0                           & 0.000592                       & 0.000636                       & 0.000679                  & 0.95                       \\
$E[Y(1)|Z=0]$           & -1.75E-05                   & 0.000102                       & 0.000104                       & 0.000123                  & 0.93                       \\
VE            & 0.004                       & 0.245                          & 0.231                          & 0.26                      & 0.93                      
\label{tab:sim_res1}
\end{tabular}
\end{table}

\subsubsection{Simulation Study 2: Non-zero bias functions and include sensitivity analysis}

The purpose of the second simulation study is to estimate power of the different methods to meet provisional approval success criteria under a realistic implementation of the methods, defined as the 95\% EUI for VE lying above 0.30.
Realistic implementation means carrying out the analysis building in conservative margin via specification of the bias functions.  Because for GBS the observational and phase 3 study populations are similar, we set $u^{UC} = 0$ and focus on the $u^{CT}$ bias function that specifies how much the surrogate endpoint $S$ departs from perfection.  In particular, following Section \ref{RoadmapStep6},
we use formula (\ref{eq: PTEspec}), setting TE = $0.7$,
$P(Y(0)=1|Z=0) = 0.005$, and PTE($X$) set to 0.67, 0.83, or 1.0, yielding
constant bias function based on the exercise to set a plausible range for $U^{CT} = U^{CT}(X,S)$
equal to 0.0012, 0.00060, or 0, respectively.  

Data are simulated under the same conditions as described in Simulation Study 1, with number sampled for $S$ measurement in each treatment arm of the $Z=0$ trial fixed at 250. We simulate data under true $U^{CT} = 0$, to study the power of the methods when they conservatively assume more bias than is present. For each simulation condition and bias setting, we report bias, standard error, confidence interval coverage, and empirical probability of provisional approval success. 

\subsubsection{Simulation Study 2: Results}

\begin{figure}[h!]
    \centering
    \includegraphics[width=0.7\textwidth]{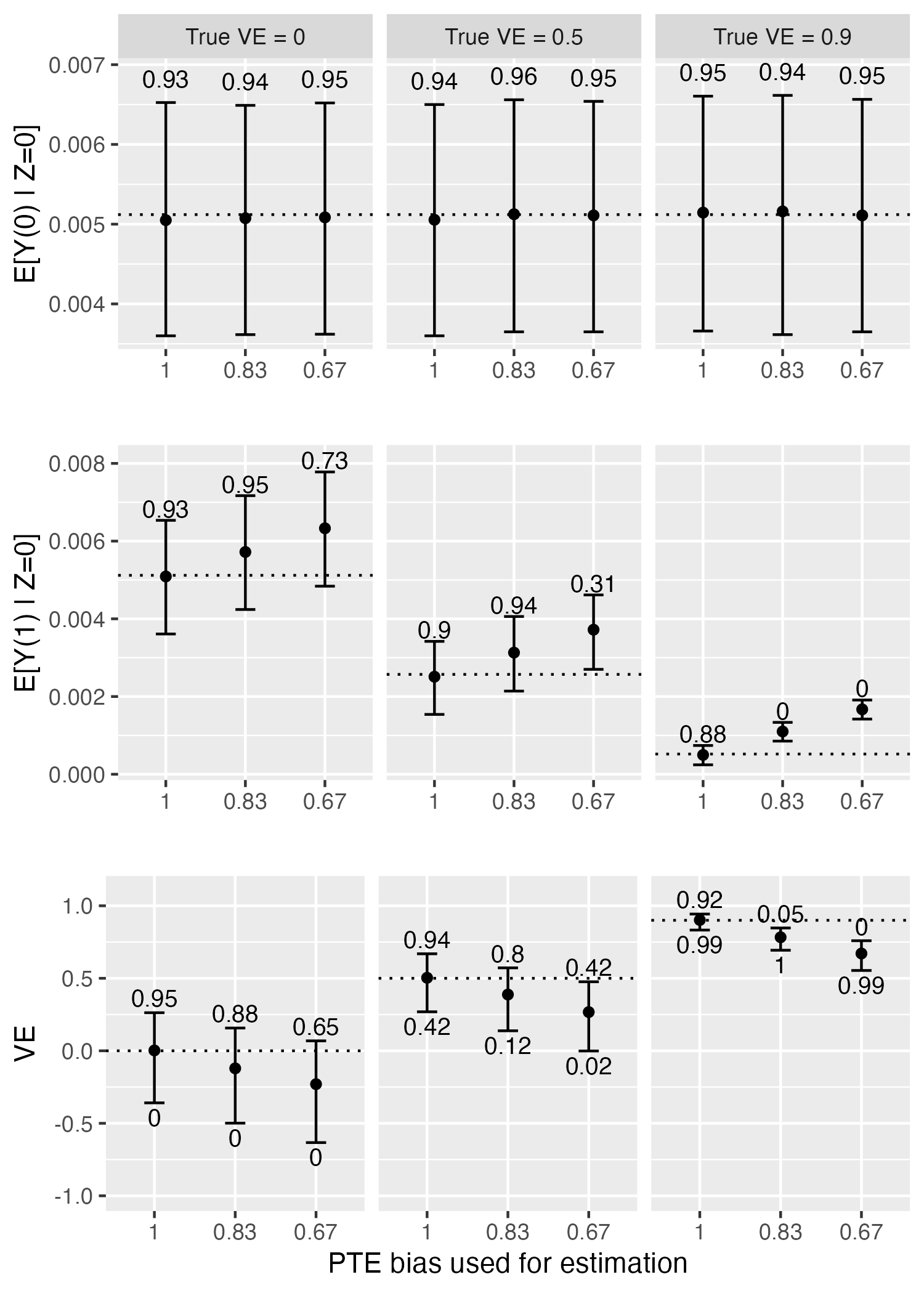}
    \caption{Results for Simulation Study 2 across true VE = 0, 0.5, and 0.9. The top, middle, and bottom panels display results for estimating $E[Y(0)|Z=0]$, $E[Y(1)|Z=0]$, and VE, respectively. Each graph depicts results across different bias functions 
    $u^{CT}(X,S) = u^{CT}$ used for estimation corresponding to PTE = 1, 0.83, and 0.67 as described in Section
    \ref{eq: PTEspec}. Points represent the median estimate across all simulations, and error bars display the median  lower and upper sandwich 95\% confidence interval bounds. Dotted horizontal lines are placed at the true value. The numbers above each error bar display the confidence interval coverage, and the numbers below the error bar in the VE panel show power to meet the success criteria (defined as 95\% EUI for VE $\ge 0.3$).}
    \label{fig:sim_res2}
\end{figure}

Results for Simulation Study 2 are shown in Figure \ref{fig:sim_res2} and Table \ref{tab:sim_res2}. We observe over-estimation of $E[Y(1)|Z=0]$ and underestimation of VE as we increase the $U^{CT}$ bias used for estimation, while estimation of $E[Y(0)|Z =0]$ is unaffected. This is as expected, since the purpose of including the $U^{CT}$ bias function is to conservatively make estimates of $E[Y(1)|Z = 0]$ larger and estimates of VE smaller. The estimator $\widehat{E}[Y(0)|Z=0]$ is unchanged because we do not vary $u^{UC}$ bias. For the true VE = 0.5 scenario, estimated empirical provisional success probabilities were 42\%, 12\%, and 2\%, for increasing specified bias based on setting PTE $= 1, 0.83$, and $0.67$, respectively. For the true VE = 0.9 scenario, estimated success probabilities were 99\%, 100\%, and 99\% for the three bias scenarios.

\begin{table}[]
\caption{Results for Simulation Study 2 on empirical bias, average standard error (sandwich and bootstrap), standard deviation of estimates, 95\% confidence interval coverage of $E[Y(0)|Z=0], E[Y(1)|Z=0]$, and VE, along with success probabilities. SE (bs) = bootstrap standard error, SE (sw) = sandwich standard error, SD = standard deviation of estimates, Cov = 95\% confidence interval coverage using sandwich standard errors, PTE = proportion of treatment explained used to generate bias functions, SP = success probability (defined as 95\% EUI for VE $\ge 0.3$) The standard errors for $\log (1-\widehat{{\rm VE}})$ are displayed for VE.}
\begin{tabular}{lllllll}
\rowcolor[HTML]{C0C0C0} 
              & {\color[HTML]{333333} Bias} & {\color[HTML]{333333} SE (bs)} & {\color[HTML]{333333} SE (sw)} & {\color[HTML]{333333} SD} & {\color[HTML]{333333} Cov} & \cellcolor[HTML]{C0C0C0}SP \\
\rowcolor[HTML]{B7B7B7} 
True VE = 0   &                             &                                &                                &                           &                            & \cellcolor[HTML]{C0C0C0}   \\
\rowcolor[HTML]{EFEFEF} 
PTE = 1       &                             &                                &                                &                           &                            & \cellcolor[HTML]{EFEFEF}   \\
$E[Y(0)|Z=0]$          & -7.00E-05                   & 0.000702                       & 0.000721                       & 0.000811                  & 0.93                       &                            \\
$E[Y(1)|Z=0]$          & -3.00E-05                   & 0.000698                       & 0.00072                        & 0.000764                  & 0.93                       &                            \\
VE            & 0.00254                     & 0.167                          & 0.156                          & 0.167                     & 0.95                       & 0                          \\
\rowcolor[HTML]{EFEFEF} 
PTE = 0.83    &                             &                                &                                &                           &                            & \cellcolor[HTML]{EFEFEF}   \\
$E[Y(0)|Z=0]$          & -4.50E-05                   & 0.000695                       & 0.000712                       & 0.000844                  & 0.94                       &                            \\
$E[Y(1)|Z=0]$          & 6.00E-04                    & 0.000721                       & 0.000742                       & 0.000813                  & 0.95                       &                            \\
VE            & -0.121                      & 0.157                          & 0.147                          & 0.16                      & 0.88                       & 0                          \\
\rowcolor[HTML]{EFEFEF} 
PTE = 0.67    &                             &                                &                                &                           &                            & \cellcolor[HTML]{EFEFEF}   \\
$E[Y(0)|Z=0]$          & -3.50E-05                   & 0.000693                       & 0.000714                       & 0.000825                  & 0.95                       &                            \\
$E[Y(1)|Z=0]$          & 0.00121                     & 0.000718                       & 0.000736                       & 0.000798                  & 0.73                       &                            \\
VE            & -0.23                       & 0.153                          & 0.144                          & 0.16                      & 0.65                       & 0                          \\
\rowcolor[HTML]{B7B7B7} 
True VE = 0.5 &                             &                                &                                &                           &                            & \cellcolor[HTML]{C0C0C0}   \\
\rowcolor[HTML]{EFEFEF} 
PTE = 1       &                             &                                &                                &                           &                            & \cellcolor[HTML]{EFEFEF}   \\
$E[Y(0)|Z=0]$          & -6.50E-05                   & 0.000708                       & 0.000721                       & 0.000786                  & 0.94                       &                            \\
$E[Y(1)|Z=0]$          & -6.00E-05                   & 0.000472                       & 0.00046                        & 0.000593                  & 0.9                        &                            \\
VE            & 0.0055                      & 0.218                          & 0.205                          & 0.24                      & 0.94                       & 0.42                       \\
\rowcolor[HTML]{EFEFEF} 
PTE = 0.83    &                             &                                &                                &                           &                            & \cellcolor[HTML]{EFEFEF}   \\
$E[Y(0)|Z=0]$          & 5.00E-06                    & 0.000703                       & 0.000724                       & 0.000787                  & 0.96                       &                            \\
$E[Y(1)|Z=0]$          & 0.00056                     & 0.000486                       & 0.000472                       & 0.000596                  & 0.94                       &                            \\
VE            & -0.11                       & 0.191                          & 0.179                          & 0.203                     & 0.8                        & 0.12                       \\
\rowcolor[HTML]{EFEFEF} 
PTE = 0.67    &                             &                                &                                &                           &                            & \cellcolor[HTML]{EFEFEF}   \\
$E[Y(0)|Z=0]$          & -1.00E-05                   & 0.000708                       & 0.000724                       & 0.000796                  & 0.95                       &                            \\
$E[Y(1)|Z=0]$          & 0.00115                     & 0.000469                       & 0.000458                       & 0.000617                  & 0.31                       &                            \\
VE            & -0.231                      & 0.176                          & 0.166                          & 0.192                     & 0.42                       & 0.02                       \\
\rowcolor[HTML]{B7B7B7} 
True VE = 0.9 &                             &                                &                                &                           &                            & \cellcolor[HTML]{C0C0C0}   \\
\rowcolor[HTML]{EFEFEF} 
PTE = 1       &                             &                                &                                &                           &                            & \cellcolor[HTML]{EFEFEF}   \\
$E[Y(0)|Z=0]$          & 2.50E-05                    & 0.00071                        & 0.00073                        & 0.000793                  & 0.95                       &                            \\
$E[Y(1)|Z=0]$          & -2.25E-05                   & 0.000118                       & 0.00012                        & 0.000177                  & 0.88                       &                            \\
VE            & 0.004                       & 0.297                          & 0.278                          & 0.344                     & 0.92                       & 0.99                       \\
\rowcolor[HTML]{EFEFEF} 
PTE = 0.83    &                             &                                &                                &                           &                            & \cellcolor[HTML]{EFEFEF}   \\
$E[Y(0)|Z=0]$         & 4.00E-05                    & 0.000733                       & 0.000743                       & 0.000773                  & 0.94                       &                            \\
$E[Y(1)|Z=0]$          & 0.00058                     & 0.000115                       & 0.000117                       & 0.000165                  & 0                          &                            \\
VE            & -0.115                      & 0.181                          & 0.176                          & 0.193                     & 0.05                       & 1                          \\
\rowcolor[HTML]{EFEFEF} 
PTE = 0.67    &                             &                                &                                &                           &                            & \cellcolor[HTML]{EFEFEF}   \\
$E[Y(0)|Z=0]$          & -1.00E-05                   & 0.000709                       & 0.000729                       & 0.000787                  & 0.95                       &                            \\
$E[Y(1)|Z=0]$          & 0.00115                     & 0.000116                       & 0.000117                       & 0.000178                  & 0                          &                            \\
VE            & -0.227                      & 0.155                          & 0.154                          & 0.166                     & 0                          & 0.99                      
\label{tab:sim_res2}
\end{tabular}
\end{table}

\subsubsection{Code availability}

The R code used for the simulations is available \href{https://github.com/jpspeng/gbs_ms_sims}{here}.

\section{Discussion}

Motivated and illustrated by contemporary Group B Streptococcus vaccine development, this article considers a context that occurs for many rare diseases, where there are promising preventive interventions and a promising surrogate endpoint(s) that is strongly associated
with a target disease outcome of interest (learned from observational studies), yet it has proven elusive to conduct a pivotal phase 3 trial that could provide direct evidence demonstrating a beneficial intervention effect to prevent the target outcome. We have proposed a statistical framework for combining prospective observational study data (e.g., case-control, case-cohort), which include data on the putative surrogate and the target outcome and can be used to estimate their relationship, with a phase 3 surrogate endpoint study that collects the same data, but, because by design it is massively under-powered to assess the treatment effect (TE) on the target outcome, the goal is conservative inference for TE based on the surrogate endpoint to support provisional approval, which would be followed by post-approval validation of direct benefit on the target outcome. Indeed, it is this under-powering that led to our approach to treat the Control-to-Treated bias function $u^{CT}$ as a specified parameter by the user instead of estimating it from the data.
This article applies the Causal Roadmap rubric to the provisional approval objective, mapping a recipe for defining target parameters, identifiability assumptions, estimators, and optimization of those estimators.  One of the needed assumptions, A3, requires a common support of the surrogate endpoint in the investigational and control arms of the phase 3 trial -- but if the new intervention is highly promising it may induce higher levels of the surrogate than attained in any control arm participants. To address this kind of challenge, consistent with the objective of the provisional approval paradigm to seek conservative/lower bound inference about TE, one approach would truncate surrogate values at the maximal value observed in control arm participants.

Our approach to estimating TE is based on transporting a regression model learned in one or more observational studies to define an estimated optimal surrogate and transport this surrogate to application in the phase 3 trial.
An alternative approach to estimating TE given the data set-up considered in this article would be based on stochastic interventional effect analysis \citep{diaz2012population,haneuse2013estimation,Hejazietal2021}, where the last citation was developed specifically to handle two-phase sampling designs (e.g., case-control, case-cohort) for measuring the intermediate outcome $S$.  In this approach,
for a univariable $S$, the method of \cite{Hejazietal2021} could be applied to the observational study to estimate how much the overall risk of disease $E[Y(0)|Z=1]$ would change under a hypothetical stochastic interventional shift upwards of $S$, where the shift is specified to match the effect of the candidate treatment vs. control on $S$ measured in an external study. This provides a way to estimate the counterfactual risk $E[Y(1)|Z=1]$ had hypothetically the treatment been administered in the observational study. Transportability assumptions like A4 and A6 for each of the treatment and control conditions would then
provide a path to estimation of  
$E[Y(0)|Z=0]$, $E[Y(1)|Z=0]$, and TE.
The required positivity assumption for this method poses a challenge (e.g., discussed in \cite{haneuse2013estimation}),
where if everyone's $S$ value is shifted upwards $\delta$ units, then there are no observed data above the maximum value of $S$ minus $\delta$ in the observational study to support estimation.  This overlapped support condition is a different kind of positivity assumption from what is required from the approach that we considered as expressed in A3.  

The transport/bridging approach considered here separates bridging into two parts: Untreated-to-Control-transport  from the observational study (of untreated individuals) to the control/placebo arm of the phase 3 study, and Control-to-Treated-transport within the phase 3 study. While we have suggested this two-part approach has advantage of aiding transparency of assumptions and study design, a potential drawback is 
the use of two sets of bridging weights that could increase variability and elevate the risk of unstable inference.
An alternative approach would use only one set of weights (e.g., 
\cite{chattopadhyay2023one}). 

\section{Acknowledgements}

The National Institute Of Allergy and Infectious Diseases of the National Institutes of Health
(R37AI054165 to P.B.G.). The content is solely the responsibility of the authors
and does not necessarily represent the official views of the National Institutes of Health, and should not be construed to represent FDA's views or policies.  The content does not represent the views of the Department of Veterans Affairs or the United States Government.

\noindent
\bibliographystyle{unsrtnat}
\bibliography{ref}

\newpage

\section{Appendix A: Mapping of causal parameters to statistical estimands}

To show identification of $E[Y(0)|Z = 0]$ in equation (\ref{eq: Untreatedtransport}), we consider the parameter within levels of baseline covariates $X$:
\begin{flalign*}
E[Y(0)|X, Z = 0] &=E[Y(0)| X , Z = 0, A = 0] &&\text{(randomization A2)} \\
&= E[E[Y(0)| X, Z = 0, A = 0, S(0)] | X, Z = 0, A=0] &&\text{(iterative expectation)} \\ 
&= E[E[Y(0)| X, Z = 1, A = 0, S(0)] | X, Z = 0, A=0] \\
& \hspace{.5in} - E[\mu^{UC}(X,S(0)) | X, Z=0, A=0] &&\text{(A4, A5)} \\
&= E[ g(X,S) | X, Z = 0, A=0] \\
& \hspace{.5in} - E[\mu^{UC}(X,S) | X, Z=0, A=0]. &&\text{(causal consistency A1)}
\end{flalign*}

\noindent By averaging $E[Y(0)|X,Z=0]$ with respect to the distribution of $X$ conditional on $Z=0$, we then obtain:  
\begin{flalign*}
E[Y(0)|Z = 0] &=E\big\{E[ g(X,S) | X, Z = 0, A=0] \mid Z=0\big\}&& \\
& \hspace{.5in} - E\big\{E[\mu^{UC}(X,S) | X, Z=0, A=0]\mid Z=0\big\}. &&  
\end{flalign*} 


\noindent Lastly, from \citet{RosevanderLaan2011},
$g(X,S):=E(Y\mid X,Z=1,A=0,S)$ and each 
$g^*_a(X, S)$
defined in Section \ref{roadmapstep5} is identified by equation
(\ref{eq: rosevan}). 

\comment{
\id{I do not think we can use MAR to get the above expression, since the MAR assumption A7 conditions on $Y$ (and I think more generally it should condition on ($\Delta$, $\tilde T$)). Instead, what we can do here is to say that $g(X,Z,S)=E(Y\mid X,Z,A=0,S)$ is identified as:
\[g=\arg\min_{\tilde g}E\left[\frac{\epsilon_S}{P(\epsilon_S=1\mid X, Z, A=0, \Delta, \tilde T)}L(\tilde g)(\Delta, \tilde T, X, Z, S)\,\bigg|\, A=0\right],\]
where $L(\tilde g)(\Delta, \tilde T, X, Z, S)$ is the loss function that we would have used if $S$ was measured for all units. This can then be fitted by simply adding weights to any regression procedure.}
}

For identification of $E[Y(1)|Z = 0]$ in equation (\ref{eq: Controltotreatedtransport}), let $H^{01}(s|x)$ be the cdf of $S(1)$ conditional on $X,Z=0,A=1$.  Calculations show:

\begin{flalign*}
&E[Y(1)|X=x,Z = 0] =E[Y(1)| X=x,Z = 0, A = 1] &&\text{(A2)} \\
&= E[E[Y(1)| X=x, Z = 0, A = 1, S(1)] | X=x,Z = 0, A=1] &&\text{(iter. exp.)} \\ 
&= \int E[Y(1)| X=x, Z = 0, A = 1, S(1)=s] dH^{01}(s|x) \\
&= \int \left\{ E[Y(0)| X=x, Z = 0, A = 0, S(0)=s] + u^{CT}(x,s) 
\right\} dH^{01}(s|x) &&\text{(A2, A3, A6) }\\
& = \int \left\{ E[Y(0)| X=x, Z = 1, A = 0, S(0)=s] + u^{CT}(x,s) - u^{UT}(x,s) 
\right\} dH^{01}(s|x) &&\text{(A4, A5) }\\
&= E[E[Y(0)|X=x, Z=1, A=0, S(0)] | X=x,Z=0,A=1]  \\
& \hspace{.05in} + E[\mu^{CT}(x,S) | X=x,Z=0,A=1] - E[\mu^{UC}(x,S) | X=x,Z=0, A=1]  \\
&= E[g(x,Z=1,S) | X=x,Z=0,A=1]  \\
& \hspace{.05in} + E[\mu^{CT}(x,S) | X=x,Z=0,A=1] - E[\mu^{UC}(x,S) | X=x,Z=0, A=1]. &&\text{(A1)} \\
\end{flalign*} 

\noindent 
The same as done for $E[Y(0)|Z=0]$, by averaging 
$E[Y(1)|X,Z = 0]$ with respect to the 
distribution of $X$ conditional on $Z=0$, we obtain:  
\begin{flalign*}
E[Y(1)|Z = 0] &= E\big\{E[ g(X,S) | X, Z = 0, A=1] \mid Z=0\big\}&& \\
& \hspace{.05in} + E\big\{E[\mu^{CT}(X,S) - \mu^{UC}(X,S)| X,  Z=0,A=1] \mid Z=0\big\}. &&  
\end{flalign*} 

\noindent Lastly, exactly as for $E[Y(0)|Z=0]$, from \citet{RosevanderLaan2011},
$g(X,S):=E(Y\mid X,Z=1,A=0,S)$ 
and each $g^*_a(X, S)$
defined in Section \ref{roadmapstep5} is identified by equation
(\ref{eq: rosevan}).

\comment{
Identification of $E[Y(0)|Z = 0]$ in equation (\ref{eq: Untreatedtransport}):

\begin{flalign*} 
E[Y(0)|Z = 0] &= E_X[E[Y(0)| X, Z = 0] | Z = 0] &&\text{(iterative expectation)} \\
&= E_X[E[Y(0)| X, Z = 0, A = 0] | Z = 0] &&\text{(A2)} \\ 
&= E_X[E[Y| X, Z = 0, A = 0] | Z = 0] &&\text{(A1)} \\
&= E_X[E_S[E[Y| X,S, Z = 0, A = 0]|X, Z = 0, A = 0] | Z = 0] &&\text{(iterative expectation)} \\
&= E_X[E_S[E[Y| X,S, Z = 1, A = 0]|X, Z = 0, A = 0] | Z = 0] \\
&\ \ \ - E_X[E_S[u^{UC}(x,s)|X, Z=0, A=0]|Z=0] &&\text{(A4)}
\end{flalign*} 

Suppose that we made an assumption of full mean exchangeability. That is $E[Y(0)|Z = 0] = E[Y(0)|Z = 0, A = 0]$. Then we have the following identification result:
\begin{flalign*} 
E[Y(0)|Z = 0] &= E[Y(0)|Z = 0, A = 0] &&\text{(full exchangeability)} \\ 
&= E[Y|Z = 0, A = 0] &&\text{(A1)} \\
&= E_{X, S}[E[Y|X, S, Z = 0, A = 0]|Z = 0, A = 0] &&\text{(iterative expectation)} \\
&= E_{X,S}[E[Y| X,S, Z = 1, A = 0]| Z = 0, A = 0] \\
&\ \ \ - E_{X,S}[u^{UC}(x,s)| Z = 0, A = 0] &&\text{(A4)}
\end{flalign*} 
}

\section{Appendix B: Violation of the Simplifying Assumption that All Enrolled Participants in Both Studies were Free of the Target Outcome through the Visit at $\tau$ for Surrogate Measurement}
\label{AppendixB}

Our development made the simplifying assumption that all enrolled participants in both studies did not experience the target outcome by 
the visit $\tau$ at which intermediate outcomes $S$ are measurable, i.e., $Y^{0-\tau}_i=0$ for all $i$.  In practice this is likely violated.  
We also made the simplifying assumption of no loss to follow-up before $\tau.$ To address this second issue, the methods are valid when applied only including participants observed to reach time $\tau$, by adding a random censoring assumption conditional on $(X,Z,A)$. 

To handle early target outcome events $Y$ before $\tau$, for applications where the surrogate is only meaningfully defined for participants reaching $\tau$ free of the target outcome, principal stratification may be an appropriate framework for extensions, 
as noted in Section \ref{ICEsec}, in which case
the causal parameters would be defined for the always-survivors (AS) principal stratum with $Y^{0-\tau}(1)=Y^{0-\tau}(0)=0$.
Under the ``equal early clinical risk" (EECR) assumption of no individual-level causal treatment effects on $Y$ by $\tau$ [$P(Y^{0-\tau}(1)=Y^{0-\tau}(0))=1$], then 
the approach to data analysis that simply excludes all early failure events works, as in this case both causal parameters 
$E[Y(a)|Z=0,Y^{0-\tau}(a)=0]$ equal 
$E[Y(a)|Z=0,Y^{0-\tau}(1)=Y^{0-\tau}(0)=0]$, for $a=0,1$.

For applications where it is not appropriate to assume EECR, an alternative assumption that achieves the same objective as EECR is that both of the following conditions hold with probability one:
\begin{enumerate}
    \item $Y(1)$ and $Y^{0-\tau}(0)$ are independent conditional on $(X, S(1))$, $Y^{0-\tau}(1)=0$ and $A=1$;
    \item $Y(0)$ and $Y^{0-\tau}(1)$ are independent conditional on $(X, S(0))$, $Y^{0-\tau}(0)=0$ and $A=0$,
\end{enumerate} 
\noindent where recall the time origin for
$Y(a) := I(T(a) \leq t_0)$ is $\tau$ and $Y^{0-\tau}(a)$ is the indicator of failure after enrollment by $\tau$ under assignment $a$, for $a=0,1$.  This assumption implies that conditioning on $Y^{0-\tau}(a)=0$ and $X$ is the same thing as conditioning on $Y^{0-\tau}(1)=Y^{0-\tau}(0)=0$ and  $X$, as noted in Shepherd et al. (2006)\nocite{Shepherdetal2006}.
For arm $A=1$, it requires that the risk of the target outcome under assignment $A=1$ is the same in the $\lbrace Y^{0-\tau}(1)=Y^{0-\tau}(0)=0 \rbrace$ AS  principal stratum as in the $\lbrace Y^{0-\tau}(1)=0,Y^{0-\tau}(0)=1 \rbrace$ principal stratum within levels of $X$ and $S(1)$ (for participants with $Y^{0-\tau}=0$). A similar interpretation applies for arm $A=0$. This assumption (e.g., for $A=1$) can be violated if experiencing the target outcome by $\tau$ under control assignment correlates with experiencing the target outcome by $\tau$ under treatment assignment, which could occur due to exposure or biological susceptibility factors not fully captured in the baseline covariates $X$.  Going beyond the assumptions considered above would require additional sensitivity analysis that would further enlarge the EUIs.

\section{Appendix C: Sandwich variance estimation for the plug-in estimator}

In this section, we describe how to calculate the sandwich variance for the plug-in estimators \citep{stefanski2002calculus}. For simplicity, we consider the no-bias function scenario, where our plug-in estimators are defined as:
\begin{eqnarray*}
\widehat\theta_{a,\mbox{\text{\tiny plug-in}}}=\frac{1}{n_{RCT}}\sum_{i=1}^nI(Z_i=0)\widehat E[ {\widehat g(X,S)} | X_i, Z_i = 0, A_i=a]
\end{eqnarray*}

To ease notation, we write $g(x, s) := g(X=x,Z=1,S=s)$ and define $\mu_a(x) := E[ g(X,S) | X = x, Z = 0, A=a]$, for $a = 0, 1$. Further, from previously defined notation, we have sampling probabilities $\pi(x,z,a,t,\delta) := P(\epsilon_s = 1|X=x, Z=z, A=a, \tilde T = t, \Delta = \delta)$.

Suppose that we have parametric models for nuisance functions $g$, $\mu_a$, and $\pi(\cdot)$, with parameters $\beta$, $\gamma_a$, and $\alpha$, respectively.  We therefore write our set of nuisance functions as $g(x, s; \beta)$, $\mu_{0}(x; \gamma_{0})$, $\mu_{1}(x; \gamma_{1})$, and $\pi(x,z,a,t,\delta; \alpha)$. Let $\hat{\beta}$, $\hat{\gamma}_0$, $\hat{\gamma}_1$ and $\hat{\alpha}$ denote estimators for these parameters obtained through estimating equations. First, let $h_{\pi}$ denote the estimating function used to obtain $\hat{\alpha}$: 
$$
0 = \sum_{i=1}^{n} h_{\pi} (\epsilon_{Si}, X_i, Z_i, A_i, \tilde T_i, \Delta_i; \alpha).
$$
Next, $h_{g}^C$, $h_{\mu_0}^C$, and $h_{\mu_1}^C$ denote the estimating functions used to obtain $\hat{\beta}$, $\hat{\gamma}_0$, and $\hat{\gamma}_1$ assuming complete data. To account for incomplete sampling of $S$, we use inverse probability sampling (IPS) weighting, which gives us our final estimating equations:

$$
0 = \sum_{i=1}^{n} h_{g}(X_i, S_i, Z_i, A_i, \epsilon_{Si}, \tilde T_i, \Delta_i; \beta, \alpha)
$$
$$
0 = \sum_{i=1}^{n} h_{\mu_0}(X_i, S_i, Z_i, A_i, \epsilon_{Si},\tilde T_i, \Delta_i ; \alpha, \beta, \gamma_0)
$$
$$
0 = \sum_{i=1}^{n} h_{\mu_1}(X_i, S_i, Z_i, A_i, \epsilon_{Si}, \tilde T_i, \Delta_i; \alpha, \beta, \gamma_1)
$$
where 
$$
h_{g}(X_i, S_i, Z_i, A_i, \epsilon_{Si}, \tilde T_i, \Delta_i; \beta, \alpha) := \frac{Z_i \epsilon_{Si} h_{g}^C(X_i, S_i, \tilde T_i, \Delta_i; \beta)}{\pi(X_i, Z_i, A_i, \tilde T_i, \Delta_i; \alpha)}
$$
$$
h_{\mu_0}( X_i, S_i, Z_i, A_i, \epsilon_{Si}, \tilde T_i, \Delta_i; \alpha, \beta, \gamma_0) := \frac{(1 - Z_i) (1-A_i) \epsilon_{Si} h_{\mu_0}^C (g(X_i, S_i, \tilde T_i, \Delta_i; \beta), X_i; \gamma_0)}{\pi(X_i, Z_i, A_i, \tilde T_i, \Delta_i; \alpha)}
$$
$$
h_{\mu_1}(X_i, S_i, Z_i, A_i, \epsilon_{Si}, \tilde T_i, \Delta_i; \alpha, \beta, \gamma_1) := \frac{(1 - Z_i) A_i \epsilon_{Si} h_{\mu_1}^C (g(X_i, S_i, \tilde T_i, \Delta_i; \beta), X_i; \gamma_1)}{\pi(X_i, Z_i, A_i, \tilde T_i, \Delta_i; \alpha)}.
$$
The plug in estimators 
$\widehat\theta_{0,\mbox{\text{\tiny plug-in}}}$ and $\widehat\theta_{1,\mbox{\text{\tiny plug-in}}}$ can be written as the solutions $\phi_0$, $\phi_1$, respectively, of the following estimating equations:
$$
0 = \sum_{i=1}^{n} (1-Z_i) \{  \mu_0(X_i; \gamma_0) - \phi_0 \} 
$$
$$
0 = \sum_{i=1}^{n} (1-Z_i) \{  \mu_1(X_i; \gamma_1) - \phi_1 \}.
$$
We define the parameter vector $\nu := (\beta, \gamma_0, \gamma_1, \alpha, \phi_0, \phi_1)$. Our final set of stacked estimating functions are 
$$
h_{\text {stack }}(X_i,S_i,Z_i,A_i,\epsilon_{Si}, \tilde T_i, \Delta_i; \nu)=\left(\begin{array}{c}h_{\pi} (\epsilon_{Si}, X_i, Z_i, A_i, \tilde T_i, \Delta_i; \alpha) \\
h_{g}(X_i, S_i, Z_i, A_i, \epsilon_{Si}, \tilde T_i, \Delta_i; \beta, \alpha) \\
h_{\mu_0}(X_i, S_i, Z_i, A_i, \epsilon_{Si}, \tilde T_i, \Delta_i; \alpha, \beta, \gamma_0) \\
h_{\mu_1}(X_i, S_i, Z_i, A_i, \epsilon_{Si}, \tilde T_i, \Delta_i ; \alpha, \beta, \gamma_1) \\
(1-Z_i) \{  \mu_0(X_i; \gamma_0) - \phi_0 \} \\
(1-Z_i) \{  \mu_1(X_i; \gamma_1) - \phi_1 \} \end{array}\right)
$$

along with the stacked estimating equation 

$$
0 = \sum_{i=1}^n h_{\text {stack }}(X_i,S_i,Z_i,A_i,\epsilon_{Si}, \tilde T_i, \Delta_i; \nu).
$$

From this estimating equation, we obtain an estimator $\hat{\nu}$. Now, we define the following:
\begin{align*}
h'_{\text {stack }}(X_i,S_i,Z_i,A_i,\epsilon_{Si}, \tilde T_i, \Delta_i; \nu) & =\partial h_{\text {stack }}(X_i,S_i,Z_i,A_i,\epsilon_{Si}, \tilde T_i, \Delta_i; \nu) / \partial \nu, \\ 
W_n\left(X_i,S_i,Z_i,A_i,\epsilon_{Si}, \tilde T_i, \Delta_i; \hat{\nu} \right) & =\frac{1}{n} \sum_{i=1}^n\left\{-h'_{\text {stack }}(X_i,S_i,Z_i,A_i,\epsilon_{Si}, \tilde T_i, \Delta_i; \hat{\nu}) \right\} \\
Q_n\left(X_i,S_i,Z_i,A_i,\epsilon_{Si}, \tilde T_i, \Delta_i; \hat{\nu} \right) & =\frac{1}{n} \sum_{i=1}^n h_{\text {stack }}(X_i,S_i,Z_i,A_i,\epsilon_{Si}, \tilde T_i, \Delta_i; \hat{\nu}) h_{\text {stack }}(X_i,S_i,Z_i,A_i,\epsilon_{Si}, \tilde T_i, \Delta_i; \hat{\nu})^T .
\end{align*}

The sandwich variance estimator for $\hat{\nu}$ is
\begin{align*}
V_n\left( X_i,S_i,Z_i,A_i,\epsilon_{Si}, \tilde T_i, \Delta_i; \hat{\nu} \right)= &W_n\left(X_i,S_i,Z_i,A_i,\epsilon_{Si}, \tilde T_i, \Delta_i; \hat{\nu}\right)^{-1} Q_n\left(X_i,S_i,Z_i,A_i,\epsilon_{Si}, \tilde T_i, \Delta_i; \hat{\nu}\right) \\ &\left\{W_n\left(X_i,S_i,Z_i,A_i,\epsilon_{Si}, \tilde T_i, \Delta_i; \hat{\nu}\right)^{-1}\right\}^T.
\end{align*}

\section{Appendix D: Simulation study details}
\label{AppendixDsimdesign}

\subsection{Design of simulation study conditions}

The simulation study was designed to roughly match published characteristics of GBS and its risk factors. 

\begin{enumerate}
    \item Probability of IGbsD by 90 days of age for colonized mothers $\approx 0.005$. From \citet{vekemans2019maternal}, approximately 20\% of mothers are colonized and 0.1\% of all infants have IGbsD. We obtain an approximate incidence of $0.005$ by assuming that only infants of colonized mothers can develop invasive GBS disease. 
    \item Geometric mean (95\% CI) of cord-blood IgG concentration 0.01 (0.01-0.02) in IGbsD cases and 0.04 (0.03 - 0.06) in controls, which was observed for infant RibN IgG in \cite{dangor2023association}.
    \item Covariates $X_1,X_2,X_3$, where $X_1$ represents pre-term birth (less than 37 weeks gestational age), 
    $X_2$ represents maternal age (younger age a risk factor), and $X_3$ represents a continuous covariate unrelated to the outcome. 
  $$  X_1 | Z, A \sim \text{Bernoulli}(0.05), X_2 | Z, A \sim \text{Uniform}(18,40),   X_3 | Z, A \sim \text{Normal}(0,1)
    $$
    \begin{itemize}
        \item Pre-term birth is associated with roughly $2.56 \times$ odds of early onset IGbsD (Puopolo et al., 2011)
        \item Maternal age $< 25$ years is associated with roughly $1.94 \times$ odds of IGbsD (Parente et al., 2017)
    \end{itemize}
\end{enumerate}

Based on these constraints, we set the following to be our true $A=0$ data generating conditional regression function: 

$$P(Y=1 | S,X_1,X_2,X_3,A=0,Z)= \beta_0+ \beta_1 S + \beta_2 X_1 + \beta_3 X_2+\beta_4 X_3
$$

where $\beta_0 = -17.1$, $\beta_1 = -8.2$, $\beta_2 = 0.69$, $\beta_3 = -0.03$, $\beta_4 = 0$, and we set our distribution of $S \sim \text{Normal}(-1.45, 0.0225)$, where $S$ represents the log IgG biomarker, $X_1 \sim \text{Bernoulli}(0.05)$, $X_2 \sim \text{Uniform}(18,40)$, and $X_3 \sim \text{Normal}(0,1)$.

With this data-generating function, we have that in our observational study (using empirical measures from a simulated $n = $10,000,000 size data set): 

\begin{itemize}
    \item Incidence of IGbsD by 90 days $\approx 0.005$
    \item Geometric mean cord-blood IgG biomarker in IGbsD cases is 0.01 and in controls is 0.04
    \item Pre-term birth associated with $2 \times$ odds of disease
    \item Each one-year increase in maternal age associated with 3\% lower odds of disease
\end{itemize}

\subsection{Results from the larger sample simulation study}
\label{simresultssupp}

In this section, we report results for a simulation study with a higher case rate.  We set the following to be our true $A=0$ data generating conditional regression function: 
$$P(Y=1 | X_1,X_2,X_3,A=0,Z,S)= \beta_0+ \beta_1 S + \beta_2 X_1 + \beta_3 X_2+\beta_4 X_3
$$
\noindent where $\beta_0 = -14$, $\beta_1 = -7$, $\beta_2 = 0.69$, $\beta_3 = -0.03$, $\beta_4 = 0$.

We preserve the same distribution of $S$ as in Simulation Study 1, with $S \sim \text{Normal}(-1.45, 0.225)$, resulting in a baseline case rate of 0.016 (about 3 times higher than 0.005).

To generate VE $= \{ 0, 0.5, 0.9\}$, we manipulate the distribution of the biomarker in the vaccine arm $S|A = 1, Z = 0$:
\begin{itemize}
    \item To create VE $= 0$, set $S|A =1, Z = 0 \sim \text{Normal}(-1.45, 0.0225)$
    \item To create VE $= 0.5$, set $S|A =1, Z = 0 \sim \text{Normal}(-1.29, 0.04)$
    \item To create VE $= 0.9$, set $S|A =1, Z = 0 \sim \text{Normal}(-1.04, 0.0441)$.
\end{itemize}

\begin{table}[]
\caption{Results for the 
Supplemental Simulation Study with approximately 3 times higher IGbsD outcome rate on
empirical bias, average standard error (sandwich and bootstrap), standard deviation of estimates, and 95\% confidence interval coverage of $E[Y(0)|Z=0], E[Y(1)|Z=0]$, and VE. SE (bs) = bootstrap standard error, SE (sw) =  sandwich standard error, SD = standard deviation of estimates, Cov = 95\% confidence interval coverage using sandwich standard error, Sampled = number with $S$ data per arm in phase 3 study. The standard errors for $\log (1-\widehat{{\rm VE}})$ are displayed for VE.}
\begin{tabular}{llllll}
\rowcolor[HTML]{C0C0C0} 
              & {\color[HTML]{333333} Bias} & {\color[HTML]{333333} SE (bs)} & {\color[HTML]{333333} SE (sw)} & {\color[HTML]{333333} SD} & {\color[HTML]{333333} Cov} \\
\rowcolor[HTML]{EFEFEF} 
True VE = 0   &                             &                                &                                &                           &                            \\
$E[Y(0)|Z=0]$ & 0                           & 0.000592                       & 0.000846                       & 0.000781                  & 0.97                       \\
$E[Y(1)|Z=0]$ & 0                           & 0.000592                       & 0.000844                       & 0.000793                  & 0.97                       \\
VE            & -0.000535                   & 0.0338                         & 0.0335                         & 0.0338                    & 0.96                       \\
\rowcolor[HTML]{EFEFEF} 
True VE = 0.5 &                             &                                &                                &                           &                            \\
$E[Y(0)|Z=0]$ & 0                           & 0.000596                       & 0.000845                       & 0.000777                  & 0.98                       \\
$E[Y(1)|Z=0]$ & -6.00E-05                   & 0.000367                       & 0.000473                       & 0.000459                  & 0.96                       \\
VE            & 0.004                       & 0.0487                         & 0.0482                         & 0.0487                    & 0.95                       \\
\rowcolor[HTML]{EFEFEF} 
True VE = 0.9 &                             &                                &                                &                           &                            \\
$E[Y(0)|Z=0]$ & 0                           & 6.00E-04                       & 0.00085                        & 0.000832                  & 0.96                       \\
$E[Y(1)|Z=0]$ & -1.00E-05                   & 0.000157                       & 0.000167                       & 0.000163                  & 0.96                       \\
VE            & 0.001                       & 0.11                           & 0.109                          & 0.108                     & 0.96                       
\label{tab:sim_res_supp}
\end{tabular}
\end{table}

Finally, we sample all 6,200 participants in the phase 3 study ($Z=1$) in this supplemental simulation study. 

Results are shown in Table \ref{tab:sim_res_supp}. In this higher case rate scenario, we observe minimal bias in estimating VE, approximately nominal confidence interval coverage, and close agreement between bootstrap, sandwich, and empirical standard errors.

\end{document}